RESEARCH ARTICLE

# Using redescription mining to relate clinical and biological characteristics of cognitively impaired and Alzheimer's disease patients

Matej Mihelčić[1,2], Goran Šimić[3], Mirjana Babić Leko[3], Nada Lavrač[2,4], Sašo Džeroski[2,4], Tomislav Šmuc[1]*, for the Alzheimer's Disease Neuroimaging Initiative[¶]

**1** Division of Electronics, Ruđer Bošković Institute, Zagreb, Croatia, **2** Jožef Stefan International Postgraduate School, Ljubljana, Slovenia, **3** Department for Neuroscience, Croatian Institute for Brain Research, University of Zagreb Medical School, Zagreb, Croatia, **4** Department of Knowledge Technologies, Jožef Stefan Institute, Ljubljana, Slovenia

¶ Membership of the Alzheimer's Disease Neuroimaging Initiative is provided in the Acknowledgments.
* tomislav.smuc@irb.hr

**Data Availability Statement:** Data containing measurements of different biological and clinical attributes for different human subjects are available from the ADNI database (http://adni.loni.usc.edu/). To obtain the data, users have to register with ADNI and provide a research statement. Currently, it is not possible to for the authors to make the data publicly available because of the ADNI data use agreement (http://adni.loni.usc.edu/wp-content/uploads/how_to_apply/ADNI_Data_Use_Agreement.pdf), point 3 that states: "I will not

## Abstract

Based on a set of subjects and a collection of attributes obtained from the Alzheimer's Disease Neuroimaging Initiative database, we used redescription mining to find interpretable rules revealing associations between those determinants that provide insights about the Alzheimer's disease (AD). We extended the CLUS-RM redescription mining algorithm to a constraint-based redescription mining (CBRM) setting, which enables several modes of targeted exploration of specific, user-constrained associations. Redescription mining enabled finding specific constructs of clinical and biological attributes that describe many groups of subjects of different size, homogeneity and levels of cognitive impairment. We confirmed some previously known findings. However, in some instances, as with the attributes: testosterone, ciliary neurotrophic factor, brain natriuretic peptide, Fas ligand, the imaging attribute Spatial Pattern of Abnormalities for Recognition of Early AD, as well as the levels of leptin and angiopoietin-2 in plasma, we corroborated previously debatable findings or provided additional information about these variables and their association with AD pathogenesis. Moreover, applying redescription mining on ADNI data resulted with the discovery of one largely unknown attribute: the Pregnancy-Associated Protein-A (PAPP-A), which we found highly associated with cognitive impairment in AD. Statistically significant correlations ($p \leq 0.01$) were found between PAPP-A and clinical tests: Alzheimer's Disease Assessment Scale, Clinical Dementia Rating Sum of Boxes, Mini Mental State Examination, etc. The high importance of this finding lies in the fact that PAPP-A is a metalloproteinase, known to cleave insulin-like growth factor binding proteins. Since it also shares similar substrates with A Disintegrin and the Metalloproteinase family of enzymes that act as $\alpha$-secretase to physiologically cleave amyloid precursor protein (APP) in the non-amyloidogenic pathway, it could be directly involved in the metabolism of APP very early during the disease course. Therefore, further studies should investigate the role of PAPP-A in the development of AD more thoroughly.

further disclose these data beyond the uses outlined in this agreement and my data use application and understand that redistribution of data in any manner is prohibited." All attribute sets used in this work are described in the appropriate electronic supplementary documents submitted with the manuscript. Short descriptions of all attributes can be seen through the Merged ADNI 1/ GO/2 Packages for R located in study info section of the download data page in the ADNI database (also described in the manuscript). Viewing the data without ADNI account is not possible, however ADNI grants access to data to any interesting researcher/phd student after registration (http://adni.loni.usc.edu/data-samples/access-data/). All evidence (redescription sets) obtained as a result of this study are available as supplementary documents submitted with the manuscript.

**Funding:** Data collection and sharing for this project was funded by the Alzheimer's Disease Neuroimaging Initiative (ADNI) (National Institutes of Health Grant U01 AG024904) and DOD ADNI (Department of Defense award number W81XWH-12-2-0012). ADNI is funded by the National Institute on Aging, the National Institute of Biomedical Imaging and Bioengineering, and through generous contributions from many private pharmaceutical companies and non-profit organizations. The research was partly funded by the Slovenian Research Agency, the Future and Emerging Technologies EU funded Human Brain Project (HBP, grant agreement no. 604102) – NL, SD, the European Commission funded MAESTRA project (Gr. no. 612944) - TS, SD, the Croatian Science Foundation funded projects: Machine Learning Algorithms for Insightful Analysis of Complex Data Structures (grant. no. 9623) - TS and Tau protein hyperphosphorylation, aggregation and trans-synaptic transfer in Alzheimer's disease: cerebrospinal fluid analysis and assessment of potential neuroprotective compounds (grant. no. IP-2014-09-9730) - GS.

**Competing interests:** The authors have declared that no competing interests exist.

## Introduction

Alzheimer's Disease (AD) is an irreversible neurodegenerative disease that results in progressive deterioration of cognitive abilities and behavioural control due to synapse and neuron loss. It is the most common cause of dementia among older adults. Although available medications for treatment of mild to moderate AD (donepezil, galantamine, and rivastigmine) and severe AD (memantine) help to some level, these drugs do not modify the underlying disease process.

The Alzheimer's Disease Neuroimaging Initiative (ADNI) [1] aims to collect various imaging and biomarker data, that could be potentially useful in diagnostics and treatment of AD. The analysis of these data provides means to potentially extend our understanding of the disease, its impact on various functions of human comportment and cognitive functions, and tracking its progression.

In this work, we analysed the data obtained from the Alzheimer's Disease Neuroimaging Initiative (ADNI) database [1], containing clinical and biological measurements (listed in S1–S3 Files and available at http://adni.loni.usc.edu/). These measurements are taken for a set of subjects in order to test for presence of AD and the level of subjects' cognitive impairment. We divided the attributes in two main groups: clinical (clin) and biological (bio).

Clinical attributes have been obtained from numerous questionnaires and neuropsychological instruments designed to test cognition and memory with the hope of early detection of AD. These tests have been carefully designed, studied and regularly updated to increase the detection of various forms of cognitive impairment. Many such tests exist [2], but there has been no unique measure that can be used to reliably make the diagnosis [3]. Thus, combining different tests has been shown to provide more reliable results. Biological attributes have contained neuroimaging data of a number of methods to visualize brain activity, such as MRI and PET scans, along with some related and derived scores. They have also contained biospecimens: a number of blood tests and measurements, and information about the subjects' genetic markers (genetic data). These attributes have been generally considered less reliable, but are still actively investigated with the aim to aid in the early detection of AD and to help understand its complex genetic, epigenetic, and environmental landscapes.

Manual investigation of associations between attributes and analysis of their effects would require insurmountable efforts, which prompted us to use a data mining technique called redescription mining.

### Work related to understanding cognitive impairment

Considerable work has been oriented towards understanding the role of biological or clinical attributes, determining correlations between different attributes and assessing their predictive power for determining the level of cognitive impairment.

Researchers have used neural imaging (MRI, PET, etc.) [4–6] to predict levels of cognitive impairment. For example, Doraiswamy et al. [7] studied PET images of subjects with cognitive decline. Donovan et al. [8] studied correlations between regional cortical thinning and worsening of apathy and hallucinations. Guo et al. [9] studied the effects of intracranial volume on association between clinical disease progression and brain atrophy or apolipoprotein E genotype. Hostage et al. [10] studied the effects of apolipoprotein E (*APOE* alleles) $\varepsilon$4 and $\varepsilon$2 on hippocampal volume. Other investigators have also studied the role of apolipoprotein E [11] in early mild cognitive impairment. These are just a few samples of the huge set of studies of correlations between biological, clinical attributes and the level of cognitive impairment. More extensive list can be found at http://adni.loni.usc.edu/news-publications/publications/.

Recently, Gamberger et al. used a multi-layer clustering method [12] to identify clusters of AD patients with respect to several clinical and biological attributes [3]. The same method was applied [13] to detect differences between clusters containing male and female patients. Breskvar et al. used Predictive Clustering Trees (PCTs) [14] to discover and analyse patient clusters. They focused on relations between biological features and the progression of AD by observing behavioural response of patients and their study partners (persons who are in frequent contact with the patient, study with the patient, and are able to assess the patient's functioning in daily life).

## Redescription mining and related fields

In this section, we provide background information related to redescription mining and motivate its choice as a data mining technique used in our work.

The most open-ended, unsupervised data-mining technique, clustering [15–19] finds and groups similar instances based on a predefined similarity measure. It is used when underlying and possibly interesting natural grouping is unavailable, but also to reveal new groups that were previously unknown. Clustering techniques typically do not create interpretable models of data, so one has to apply other technique in order to get interpretable descriptions of induced clustering. One such approach, limited to using a single attribute set, is conceptual clustering [20, 21] that aims at finding clusters that can be described with concepts derived by using some description language.

There exists a broad group of descriptive pattern mining techniques that find and describe subsets of examples using single attribute set or view.

For example, association rule mining [22] finds associations between items (in transaction databases) or different attributes in the form of unidirectional rules. Interesting associations are typically selected based on support and confidence scores of association rules and possibly some other interestingness measures.

Subgroup discovery [23, 24] is a technique that finds queries describing groups of instances having unusual and interesting statistical properties with respect to the target variable. Contrast Set Mining [25] identifies monotone conjunctive queries that best discriminate between instances containing one target class from all other instances (e.g. subjects with diagnosis Alzheimer's Disease (AD) vs Control (CN) subjects).

In contrast to techniques operating on a single set of attributes, multi-view techniques offer advantages when the available data contains information from various sources or descriptions of different properties of instances (as is the case in this study).

Two-view data association discovery [26] aims at finding a small, non—redundant set of associations that provide insight in how two views are related. The approach can create both *bidirectional* and *unidirectional* rules as translation patterns.

Redescription mining, introduced by Ramakrishnan et al. [27], is capable of mining descriptions of subsets of data described by multiple sets of attributes. The building blocks of redescriptions are called queries (logical formulas describing a set of instances by using attributes from some particular view). Redescription queries can describe the same or very similar subset of instances with different queries, which is an important capability in the context of knowledge discovery.

## Rationale for using redescription mining

Redescription mining offers advantages over related techniques and provides specific results required for our analysis. The multi-view and descriptive capabilities of redescription mining make it suitable for relating different biological attributes, many with unknown or scarcely explored role and effects on cognitive impairment, to clinical attributes designed to detect cognitive impairment and make the preliminary diagnosis.

Although a two-view data association discovery approach can be applied to this data, we aimed at discovering interesting equivalence-like associations between biological and clinical attributes on different support levels and validating them with the subjects diagnosis, that is possible with redescription mining. Two-view association discovery is also somewhat limited as it is designed to mine Boolean data and to provide small and non-redundant sets of associations (translations) between different attribute sets. In our discovery study we aim to create, potentially larger number, of understandable redescriptions that would be used as a basis for the thorough statistical analyses and the analysis performed by the domain expert.

Similar data and attributes, related to AD, have been studied before [3, 13, 14, 28]. However, this study is focussed on the analysis of the ADNI data using redescription mining, which enables using its specific advantages over other approaches to find potentially new insights and improve our understanding of the genesis of AD.

## Materials and methods

This section contains descriptions of data, notation and related redescription mining approaches, CLUS-RM algorithm [29, 30] and the motivation for its use in this work. It includes description of algorithmic extensions incorporated into CLUS-RM that enable fully automated constraint-based redescription mining, where we generalize the attribute and instance level constraints introduced by Zaki and Ramakrishnan [31].

### Data description

For this study, we extracted data from the ADNI database [1]. To obtain the data, we used the *Merged ADNI 1/GO/2 Packages for R* [32] located in study info section of the download data page in the database. This package contains majority of available datasets in the format of R data frames. The basis of our datasets was contained in the adnimerge data table, which contains measurement of several clinical attributes (derived by using questionnaires, observations by doctors and other tests measuring level of cognition) and biological attributes (different blood tests, genetic markers, attributes derived from brain images, volumes of different parts of the brain etc.) for 1,737 subjects. There was also a target variable—diagnosis (not used for redescription construction) containing categorical values: control normal (CN), significant memory concern (SMC), early mild cognitive impairment (EMCI), late mild cognitive impairment (LMCI) and probable AD. Values of a target variable can be considered as ordered (levels of cognitive impairment). Each subject was assigned in exactly one category and there were no missing values for this variable. By examining the subjects contained in the adnimerge data table, we have noticed two distinct groups of subjects for whom some additional distinct attributes were measured. Therefore, we created and studied three related datasets.

The distributions of patients, divided by the level of cognitive impairment, for all three datasets are provided in Table 1.

Division of attributes to clinical (clin) and biological (bio) forms two disjoint sets of attributes used as views in redescription mining. In all datasets, subjects or patients constitute the instances for the redescription mining process.

Table 2 contains full names and abbreviations for all attributes required to present our work, while Tables 3 and 4 contain corresponding basic statistical information for these attributes. Due to data normalization (especially of biological attributes), the original measuring units do not correspond to the attribute values and are not specified in the tables.

Table 1. The number of subjects contained in datasets $D_1$, $D_2$ and $D_3$ divided by the level of cognitive impairment.

| Dataset | Total | CN  | SMC | EMCI | LMCI | AD  |
|---------|-------|-----|-----|------|------|-----|
| $D_1$   | 1737  | 417 | 106 | 310  | 562  | 342 |
| $D_2$   | 918   | 188 | 106 | 310  | 164  | 150 |
| $D_3$   | 820   | 229 | 0   | 1    | 398  | 193 |

https://doi.org/10.1371/journal.pone.0187364.t001

Table 2. A list of clinical and biological attributes discussed in the text.

| Attribute (bio) | Full name | Attribute (bio) | Full name |
|---|---|---|---|
| $A\beta_{1-40}$ | Plasma biomarker $A\beta_{1-40}$ | ICV | Intracranial volume |
| $A\beta_{1-42}$ | Plasma biomarker $A\beta_{1-42}$ | Insulin | Insulin |
| ANG2 | Angiopoietin-2 | Leptin | Leptin |
| APAII | Apolipoprotein A-II | MCRPHMIF | Macrophage migration inhibitory factor |
| APOB | Apolipoprotein B | PAPP-A | Pregnancy associated plasma protein A/ pappalysin-1 |
| APOE $\varepsilon 4$ | Gene APOE $\varepsilon 4$ | PLMNRARC | Pulmonary and activation-regulated chemo |
| AV45 | $^{18}$F-florbetapir | PPP | Pancreatic polypeptide |
| BAT126 | Level of vitamin B12 | PTAU | Phospho-tau protein |
| BNP | Brain natriuretic peptide | RCT11 | Serum glucose |
| CKMB | Creatine kinase level | RCT12 | Total protein |
| CNTF | Ciliary neurotrophic factor | RCT14 | Creatine kinase |
| Entorhinal | Entorhinal cortex volume | SPARE_AD | Spatial Pattern of Abnormalities for Recognition of Early AD |
| FASL | Fas ligand | T2TCV | T2 weighted total intracranial volume |
| FDG-PET | $^{18}$fluorodeoxyglucose—positron emission tomography | TAU | Tau protein |
| Fusiform | Volume of the fusiform gyrus | TNC | Tenascin-C |
| Hippocampus | Hippocampus volume | TSTSTRNT | Total blood testosterone |
| HMT8 | Neutrophils | Ventricles | Volume of the lateral ventricles |
| HMT18 | Eosinophils | WholeBrain | Whole brain volume |
| Attribute (clin) | Full name | Attribute (clin) | Full name |
| ADAS11 | 11-item ADAS test score | CDRSB | Clinical Dementia Rating Sum of Boxes |
| ADAS13 | 13-item ADAS test score | EcogPtPlan | Participant everyday cognition planning |
| BCNAUSEA | Presence of nausea | FAQ | Functional Assessment Questionnaire |
| BCSWEATN | Presence of sweating | MMSE | Mini-Mental State Examination |
| BCVOMIT | Presence of vomiting | MOCA | Montreal Cognitive Assessment |
| CDGLOBAL | Global cognitive dementia rating | Q13SCORE | Question 13 from the ADAS test |
| CDJUDGE | Judgement and problem solving score | RAVLT | Rey Auditory Verbal Learning Test immediate |
| CDMEMORY | Memory score | | |

https://doi.org/10.1371/journal.pone.0187364.t002

The first dataset ($D_1$) contained 1,737 subjects. The dataset contained a number of biological attributes such as *APOE* genotype, different brain measurements, such as the volume of the whole brain, the hippocampus, ventricles, and many other structures, including brain images obtained by using the $^{18}$fluorodeoxyglucose (FDG)-PET method. The dataset contained various blood analysis, such as levels of white and red blood cells, protein (RCT12) and glucose (RCT11) levels, and many others. It also contained a number of neuropsychological tests, such as the Alzheimer Disease Assessment Scale (ADAS11, ADAS13, etc.), several different Rey Auditory Verbal Learning Tests (RAVLT), Mini-Mental State Examination (MMSE), Functional Assessment Questionnaire (FAQ), and others, including several attributes related to clinical dementia rating (CDR) and geriatric depression scale (GDS). Several features describing the subject's symptoms, such as presence of nausea (BCNAUSEA), vomiting (BCVOMIT),

**Table 3. Information about value range and percentage of missing values for biological attributes discussed in the text.** Absence of an attribute from a dataset is denoted with "-" in the range and missing columns.

| Attribute | $D_1$ | | $D_2$ | | $D_3$ | |
|---|---|---|---|---|---|---|
| | Range | Missing | Range | Missing | Range | Missing |
| APOE ε4 | {0, 1, 2} | 1% | {0, 1, 2} | 2% | {0, 1, 2} | 0% |
| BAT126 | [96, 6725] | 12% | [96, 6725] | 15% | [99, 3429] | 8% |
| Entorhinal | [1426, 5896] | 16% | [1438, 5896] | 13% | [1426, 5731] | 39% |
| Fusiform | [8991, 29950] | 16% | [10012, 29950] | 13% | [8991, 24788] | 39% |
| Hippocampus | [2991, 10769] | 14% | [2991, 10602] | 10% | [3091, 10769] | 19% |
| HMT8 | [0.98, 11.64] | 12% | [1.22, 10.22] | 15% | [0.98, 11.64] | 7% |
| HMT18 | [0, 35.8] | 12% | [0, 24] | 15% | [0, 34.8] | 7% |
| ICV | $[1.1, 2.1] \cdot 10^6$ | 1% | $[1.1, 2.1] \cdot 10^6$ | 2% | $[1.1, 2.1] \cdot 10^6$ | 0% |
| RCT11 | [55, 413] | 11% | [61, 315] | 15% | [55, 413] | 6% |
| RCT12 | [5.7, 9.7] | 11% | [5.9, 8.4] | 15% | [5.7, 9.7] | 6% |
| RCT14 | [18, 2658] | 11% | [23, 2658] | 15% | [18, 721] | 6% |
| Ventricles | $[0.6, 1.5] \cdot 10^5$ | 5% | $[0.6, 1.3] \cdot 10^5$ | 7% | $[0.6, 1.5] \cdot 10^5$ | 2% |
| WholeBrain | $[0.7, 1.5] \cdot 10^7$ | 3% | $[0.8, 1.5] \cdot 10^7$ | 4% | $[0.7, 1.4] \cdot 10^7$ | 1% |
| AV45 | [0.84, 2.03] | 49% | [0.84, 2.03] | 3% | - | - |
| FDG-PET | [3.49, 8.54] | 25% | [3.49, 8.54] | 2% | - | - |
| PTAU | - | - | [9.4, 173.3] | 58% | - | - |
| $A\beta_{1-40}$ | - | - | - | - | [13.0, 371.8] | 13% |
| $A\beta_{1-42}$ | - | - | - | - | [4.6, 102.8] | 12% |
| ANG2 | - | - | - | - | [0.11, 1.46] | 31% |
| APOAII | - | - | - | - | [2.35, 3.18] | 31% |
| APOB | - | - | - | - | [2.89, 3.47] | 31% |
| BNP | - | - | - | - | [1.86, 4.13] | 31% |
| CKMB | - | - | - | - | [−1.43, 0.59] | 31% |
| CNTF | - | - | - | - | [0.88, 3.48] | 31% |
| FASL | - | - | - | - | [0.85, 3.62] | 31% |
| Insulin | - | - | - | - | [−0.68, 1.43] | 31% |
| Leptin | - | - | - | - | [−0.82, 2.0] | 31% |
| MCRPHMIF | - | - | - | - | [−1.2, 0.8] | 31% |
| PAPP-A | - | - | - | - | [−2.34, −0.85] | 31% |
| PLMNRARC | - | - | - | - | [1.6, 2.7] | 31% |
| PPP | - | - | - | - | [−0.004, 3.16] | 31% |
| SPARE_AD | - | - | - | - | [−3.86, 2.79] | 0% |
| T2TCV | - | - | - | - | [1003, 1922] | 1% |
| TAU | - | - | - | - | [19.9, 300.5] | 58% |
| TNC | - | - | - | - | [1.9, 3.5] | 31% |
| TSTSTRNT | - | - | - | - | [−1.44, 1.52] | 31% |

https://doi.org/10.1371/journal.pone.0187364.t003

sweating (BCSWEATN), as well as results of various neurological examinations were also included. Information about attributes and subjects contained in $D_1$ are available in S1 File.

The second dataset ($D_2$) contained 918 subjects. In addition to features contained in the first dataset, it also contained features describing subjects' performance on Montreal Cognitive Assessment (MOCA) scale and features related to the Eastern Cooperative Oncology Group (ECOG) Scale of Performance Status. It also contained values of cerebrospinal fluid (CSF), total tau (TAU) and phospho-tau (PTAU) levels. Information about attributes and subjects contained in $D_2$ are available in S2 File.

Table 4. Information about value range and percentage of missing values for clinical attributes discussed in the text. Absence of an attribute from a dataset is denoted with "-" in the range and missing columns. If some dataset has equal range as $D_1$, this is denoted with "-||-" in the appropriate field.

| Attribute | $D_1$ | | $D_2$ | | $D_3$ | |
|---|---|---|---|---|---|---|
| | Range | Missing | Range | Missing | Range | Missing |
| ADAS11 | [0, 42.67] | 0% | [0, 40] | 0% | [0, 40] | 0% |
| ADAS13 | [0, 54.67] | 1% | [0, 52] | 1% | [0, 52] | 1% |
| BCNAUSEA | {0, 1} | 0% | -||- | 0% | -||- | 0% |
| BCSWEATN | {0, 1} | 0% | -||- | 0% | -||- | 0% |
| BCVOMIT | {0, 1} | 0% | -||- | 0% | -||- | 0% |
| CDGLOBAL | {0, 0.5, …2} | 0% | -||- | 0% | {0, 0.5, 1} | 0% |
| CDJUDGE | {0, 0.5, …, 3} | 0% | -||- | 0% | -||- | 0% |
| CDMEMORY | {0, 0.5, …, 3} | 0% | -||- | 0% | {0, …, 2} | 0% |
| CDRSB | {0, 0.5, …, 10} | 0% | -||- | 0% | {0, …, 9} | 0% |
| FAQ | {0, 1, …, 30} | 1% | {0, 1, …, 28} | 1% | -||- | 0% |
| MMSE | {18, 19, …, 30} | 0% | {19, …, 30} | 0% | -||- | 0% |
| Q13SCORE | {0, 0.5, …, 10} | 1% | -||- | 0% | -||- | 1% |
| RAVLT | {0, 1, …, 71} | 0% | {1, …, 71} | 0% | {0, …, 69} | 0% |
| EcogPtPlan | - | - | [1, 4] | 1% | - | - |
| MOCA | - | - | {4, 5, …, 30} | 1% | - | - |

https://doi.org/10.1371/journal.pone.0187364.t004

The third dataset ($D_3$) contained 820 subjects. It was extremely useful to study the differences and special properties of healthy subjects as compared to patients with severe stages of dementia. This dataset lacked information about ECOG Scale of Performance Status, MOCA, and information about CSF biomarkers, but it contained several additional attributes related to hormones and proteins measured. It also contained information about T2 weighted total cranial vault segmentation (T2TCV) and plasma biomarkers $A\beta_{1-40}$ and $A\beta_{1-42}$. One particularly useful imaging attribute was Spatial Pattern of Abnormalities for recognition of early AD (SPARE_AD), which was specifically constructed to help in early detection of AD. Dataset $D_3$ also contained the attribute PAPP-A which is analysed in more detail in this work. The AD assessment scale contained many additional attributes corresponding to different cognitive tasks, the full set of attributes being publicly available on the ADNI web page http://adni.loni.usc.edu/. Information about attributes and subjects contained in $D_3$ are available in S3 File.

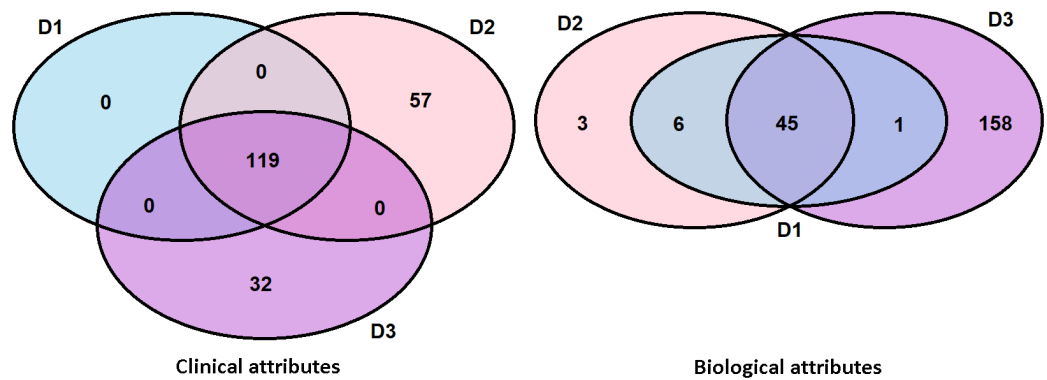

Fig 1. Relations between attributes used in constructed datasets $D_1$, $D_2$ and $D_3$. Left Venn diagram depicts clinical and right Venn diagram biological attributes.

https://doi.org/10.1371/journal.pone.0187364.g001

Relation between attributes used in different datasets is visible in Fig 1.

Division among subjects in the constructed datasets is as follows: $D_1 = D_2 \cup D_3$, $D_2 \cap D_3 = \{2002\}$, where 2002 denotes the roster id (RID), unique id of subject contained in the intersection.

In all analysed datasets, there were slightly more males than females. Males constitute 55% of the first, 52% of the second and 58% of the third dataset. They also constitute 57%, 53% and 61% of all subjects with some level of cognitive impairment in these datasets. Pregnancy in female subjects can alter levels of PAPP-A attribute. Although the information about the pregnancy status for female subjects analysed was not directly available in our dataset, documents describing ADNI1 exclusion criteria (which cover patients contained in our dataset $D_3$) [33] clearly state that female participants must be sterile or two years past childbearing potential to be included in the study group. Documents related to ADNIGO [34] and ADNI2 exclusion criteria [35] state that the participant must not be pregnant, lactating or of childbearing potential. As a result of these exclusion criteria, we can assume that the PAPP-A levels, for the studied female subjects, were not influenced by pregnancy.

## Redescription mining

Redescription mining [27] works on a dataset $D$, containing $|D|$ instances and one set, or two disjoint sets of attributes (views, denoted as $W_1$ and $W_2$) describing these instances. A redescription (as for example $R = (q_1, q_2)$) is a pair of queries, containing one query per view. Each query is a propositional logic formula that can contain conjunction, disjunction or negation operators and is used to define conditions on values of a subset of attributes from a particular view. The subset of instances described by a query $q_i$, denoted $supp(q_i)$ is called the query support set. The support set of a redescription is the set of instances described by both queries that constitute this redescription: $supp(R) = supp(q_1) \cap supp(q_2)$. We also use the notation $E_{1,1}$ to denote the set of instances described by both queries, $E_{1,0}$ a set of instances described by the first query but not described by the second query, $E_{0,1}$ a set of instances described by the second query but not described by the first query, $E_{0,0}$ a set of instances that are not described by either query. $E_{?,1}$ denotes a set of instances for which it is not possible to determine if they are described by the first query, due to missing values, but are described by the second query, $E_{1,?}$ contains a set of instances described by the first query but for which it is not possible to determine if they are described by the second query, due to missing values, $E_{?,0}$ denotes a set of instances for which it is not possible to determine if they are described by the first query, due to missing values, and are not described by the second query, $E_{0,?}$ contains a set of instances not described by the first query but for which it is not possible to determine if they are described by the second query, due to missing values. The set $E_{?,?}$ contains instances for which it is not possible to determine if they are described by either query due to missing values. $attr(R)$ denotes a multiset of attributes contained in redescription queries, whereas $attrs(R)$ represents a corresponding set of attributes. $attr(D)$ denotes all attributes contained in both views of the dataset and $\mathcal{R}$ denotes a redescription set.

We evaluate the quality of mined redescriptions by using two measures [36]: i) the Jaccard index, which measures the similarity of support sets of the two redescription queries (also often called accuracy of redescription, since it measures how close two query support sets are to containing identical set of instances) and ii) statistical significance of the observed redescription, expressed through a $p$-value.

The Jaccard index is defined as:

$$J(R) = \frac{|supp(q_1) \cap supp(q_2)|}{|supp(q_1) \cup supp(q_2)|}$$

Assessment of the statistical significance of the redescription $R = (q_1, q_2)$ is based on an assumption that the support sets, of two queries $q_1$ and $q_2$, are selected randomly, with marginal probabilities $p_1 = \frac{|supp(q_1)|}{|D|}$ and $p_2 = \frac{|supp(q_2)|}{|D|}$ respectively. The statistical significance of redescription measures how probable it is to obtain overlap of the size $|supp(R)|$ or larger when sampling two subsets of instances from a set of size $|D|$, using sampling probabilities $p_1$ and $p_2$ respectively. The size of the intersection follows a binomial distribution and the probability we are looking for can hence be written as:

$$pV(R) = \sum_{n=|supp(R)|}^{|D|} \binom{|D|}{n} (p_1 \cdot p_2)^n \cdot (1 - p_1 \cdot p_2)^{|D|-n}$$

**Example 1**. Redescription $R_{ex} = (q_{clin}, q_{bio})$, discovered on dataset $D_3$, whose queries are defined as: $q_{clin}$: $0.0 \leq GDTOTAL \leq 2.0 \wedge GDALIVE = 0.0 \wedge CDMEMORY = 0.0$ $q_{bio}$: $0.5 \leq HMT18 \leq 16.0 \wedge -3.86 \leq SPARE\_AD \leq -0.93$, provides alternative descriptions of 156 different normal control subjects. Query $q_{clin}$ describes 204 subjects with specific value for the following clinical attributes: memory score (CDMEMORY), total score in geriatric depression scale (GDTOTAL), score on a question *Do you think its wonderful to be alive now?* (GDALIVE) while query $q_{bio}$ describes 172 subjects having specific values for biological attributes such as percentage of Eosinophils (HMT18) and a Spatial Pattern of Abnormalities for Recognition of Early Alzheimer's disease (SPARE_AD). The set of subjects described by at least one query of redescription $R_{ex}$ contains 220 subjects, i.e $|supp(q_{clin}) \cup supp(q_{bio})| = 220$. For 156 of 220 subjects, both queries are valid, i.e. $|supp(q_{clin}) \cap supp(q_{bio})| = 156$. This means that the Jaccard index (accuracy) for this redescription is $\frac{156}{220} = 0.709$. The redescription is statistically significant with the $p$-value $< 2 \cdot 10^{-17}$ (which can be computed by using the formula above). It means that it is highly unlikely to observe a redescription of support size 156 or larger given that we combine two statistically independent queries, with marginal probabilities $p_1 = \frac{204}{820} = 0.25$ and $p_2 = \frac{172}{820} = 0.21$, into a redescription $R_{ex}$.

**Existing approaches for redescription mining.** The first algorithm for redescription mining, called CARTwheels, was developed by Ramakrishnan et al. [27]. Several redescription mining algorithms have been developed since, all of which can handle Boolean attributes. From these, some algorithms [29, 30, 37, 38] work also with categorical and numerical attributes. Currently, only two redescription mining algorithms ReReMi [37] and CLUS-RM [29, 30], work with attributes containing missing values.

Redescription mining algorithms can be divided into three main categories: a) algorithms based on itemset mining, b) greedy algorithms and c) tree-based algorithms.

Itemset mining based redescription mining algorithms utilize different itemset mining methods to create itemsets, which are used to create redescriptions. Approach by Zaki and Ramakrishnan [31] and the approach by Parida and Ramakrishnan [39], use a lattice (partially ordered set) of attribute sets to find redescriptions. Approach developed by Gallo et al. [40] is based on frequent itemset mining. The field is known as Frequent Itemset Mining, because the notion of frequency (support size, the apriori principle) is central in obtaining practical algorithms.

Greedy algorithms for redescription mining work by incrementally updating queries with the goal of increasing redescription accuracy. The first algorithm developed in this category was the greedy algorithm from Gallo et al. [40]. This algorithm has been extended by Galbrun and Miettinen [37], under the name ReReMi, to work with categorical and numerical attributes.

Tree-based algorithms use decision trees [41] or Predictive Clustering trees (PCTs) [42] to create redescriptions. This category includes the first developed algorithm for redescription mining called CARTwheels, developed by Ramakrishnan et al. [27]. This algorithm works by building two decision trees per iteration (one for each view) that are joined in the leaves. Redescriptions are created by reading off the conditions along the paths from the root node of the first tree to some specified class (which constitutes one redescription query) and the paths from the root node to the matching leafs of the second tree (which constitutes the second redescription query). All created trees are of the same predefined depth, and the process iterates for a predefined number of iterations. This algorithm uses multiclass classification to guide the search between the two views. Layered trees (LayeredT) and Split trees (SplitT) algorithms developed by Zinchenko [38] use a different methodology of decision tree construction to obtain redescriptions. Instead of creating fully grown trees of predefined depth, the Layered trees algorithm creates one or more depth one trees at each algorithm step. For each leaf of the tree under construction, at some fixed iteration, the Layered trees algorithm builds a new depth one tree and appends it to the corresponding leaf of the existing tree (thus increasing its complexity and size). The algorithm allows creating more informed splits, since at a certain step of tree construction, the algorithm uses information about splits created at a corresponding level of the tree constructed on the opposite view. To construct a tree of maximal depth, the algorithm considers all nodes of the tree created on the opposite view (not just the leaves of a fully grown tree as in CARTwheels). The Split trees algorithm creates decision trees of increasing size. At each step of tree construction, the depth is increased by one and a whole new tree of larger depth is built (completely replacing the previously constructed tree) until trees of maximally allowed depth are built. This algorithm simultaneously refines classes (since it obtains finer splits with trees of larger depth) and trees (by increasing their complexity and providing more specific classes).

The CLUS-RM algorithm developed by Mihelcic et al. [29, 30] uses multi-target Predictive Clustering trees (PCTs) [42, 43], instead of decision trees to construct redescriptions. Using multi-target PCTs allows using information about all nodes (intermediate nodes as well as leaves) in the constructed PCT simultaneously to create redescriptions (which increases accuracy, diversity and number of produced redescriptions). This algorithm has been extended by Mihelcic et al. [44] to use a random forest of PCTs which further increases accuracy and diversity of produced redescriptions. The CLUS-RM is also equipped with a redescription set construction procedure called redescription set optimization [29, 30, 44]. It enables incorporating quality constraints in multi-objective optimization manner and uses all produced redescriptions to create a reduced redescription set of user-defined size. A generalized version of redescription set optimization has been presented by Mihelcic et al. [45]. In addition to its main purpose of redescription set construction, this procedure allows for use of ensembles of redescription mining algorithms, influencing the structure of produced sets through user-defined importance weights and performing computationally efficient construction of multiple redescription sets with different properties, which is beneficial for exploration [45].

### Choice of methodology, redescription accuracy measure and a query language

In this section, we describe our motivation underlying the use of CLUS-RM algorithm and the extensions made to allow performing constraint-based redescription mining. In addition, we describe what reasons motivated us for the use of a redescription accuracy evaluation measure and a specific query language used to construct redescriptions.

**Choice of redescription mining algorithm.** To create redescriptions, we used the CLUS-RM algorithm [29, 30] based on Predictive Clustering trees (PCT) [42, 43]. PCTs allow clustering on both target and descriptive space. By using their multi-label and multi-target capability one can use multiple (or all) nodes in a given tree simultaneously to produce redescriptions. Due to the property of inductive transfer [46], multi-target classification can outperform single-target classification, which improves the overall accuracy of produced redescriptions. The CLUS-RM algorithm incorporates a redescription set optimization procedure (a novelty compared to other redescription mining approaches), which uses the large number of diverse redescriptions produced to optimize a redescription set of user-define size.

Using a large number of produced redescriptions in the optimization process increases the quality of the redescription set presented to the user. The optimization process evaluates redescriptions according to accuracy, significance and redundancy (with respect to redescription support sets and attributes contained in redescription queries).

Since our data contain missing values, we could only use the CLUS-RM or the ReReMi algorithm to find redescriptions. Given our goal of using the produced redescription sets to perform further statistical analysis, there are several reasons that motivate the use of CLUS-RM as the redescription mining algorithm in this work. CLUS-RM has the ability to produce potentially large sets of redescriptions that can be used to perform statistical analysis (e.g. of obtained associations). Multiple different redescriptions containing the same attribute pair and describing different subsets of instances reinforce the importance of frequently co-occurring attributes. CLUS-RM can constrain redescription support set size to an interval, which provides experts with a range of associations (hypotheses), from general (intervals containing larger support set size) to more specific (intervals containing smaller support set size). It can also produce redescription sets of user defined size which allows creating sets that contain equal number of members per support interval for further statistical analysis. Because of this, association statistics will not be constructed only from very general or very specific redescriptions, but from redescriptions covering a whole range of different support sizes. The experiments with CLUS-RM [30], and its extension [44], as well as the integration of the CLUS-RM into a redescription mining framework for redescription set construction [45], show that the produced redescription sets were fully competitive with other state-of-the-art solutions, and in some cases (as when only conjunctions are used in redescription query construction), the resulting redescription sets can even contain significantly more accurate and diverse redescriptions.

To obtain the results presented in this work, we required the constraint-based redescription mining capability, mostly using one attribute as constraint. However, developing a constraint-based methodology that is able to use multiple attributes (instances) as constraint was straightforward and is also presented as a part of this work. The proposed extensions include several modes of constraint-based redescription mining (CBRM) that allow exploring interactions of multiple attributes from different views with Boolean, categorical and numerical variables, extending the state-of-the-art in CBRM. Instance level constraints can be incorporated in analogous fashion.

The one-attribute CBRM capability of Siren [47] allows selecting one attribute as constraint and defining its numerical interval (for numerical attributes). The resulting redescription set is comprised of redescriptions that are obtained by extending the initial query supplied by the user. When compared to this limited CBRM capability of Siren, the CLUS-RM extension operates in a fully automated constraint-based setting (allowing multiple attributes as constraints). Also, it is not necessary to manually select numerical bounds as is currently the case in Siren. In general, performing interactive constraint-based redescription mining can demand significant effort and time from the domain expert (in addition to examination of computed redescriptions, which also needs to be done in our approach), but can potentially enable tuning the algorithm better to find information about some specific, targeted problem.

Analysis and exploration of precomputed redescription sets, based on multiple different redescription criteria, exploration of different attribute associations and groupings of instances based on a produced redescription set is also possible with the tool InterSet [48].

**Choice of redescription accuracy measure.** Since the data contains missing values, we used the query non-missing Jaccard index, introduced in [30], and further explained in [45] to evaluate redescriptions. The query non missing Jaccard index is defined as:

$$J_{qnm}(q_1, q_2) = \frac{|E_{1,1}|}{|E_{1,1}| + |E_{?,1}| + |E_{1,?}| + |E_{0,1}| + |E_{1,0}|}$$

Query non-missing Jaccard index evaluates instances as being a part of redescription support set only if there is enough information in the data (given the query language) to deduce that these instances satisfy the conditions of both redescription queries. The construction of this measure is guided by the principle that the query cannot contain an instance in its support set if it cannot be evaluated due to missing values. Because of this, the measure does not penalize the score with instances contained in the sets $E_{?,?}$, $E_{0,?}$, $E_{?,0}$ and rather treats them as if they were contained in the set $E_{0,0}$ but penalizes the score with instances contained in the sets $E_{?,1}$ and $E_{1,?}$ and treats them as if they are contained in sets $E_{1,0}$ and $E_{0,1}$.

Query non-missing Jaccard index has been designed to trade-off between the pessimistic and the optimistic Jaccard index [36], which are each forcing opposite extreme values and are thus leading to less realistic estimates of the true Jaccard index. Query non-missing Jaccard index is optimistic because it does not penalize the score with instances that are not described by one query and cannot be evaluated by the other query, due to missing values ($E_{?,0}$, $E_{0,?}$). On the other hand, it is pessimistic, since it penalizes the score with instances that are described by one redescription query, but cannot be evaluated by the other, due to missing values ($E_{1,?}$ and $E_{?,1}$). Redescription accuracy estimates provided by query non-missing, pessimistic and optimistic Jaccard index have already been compared experimentally in [45].

**Choice of a query language.** In this work, our redescriptions consist exclusively of conjunctive queries. Queries containing only conjunction operators are easier to understand and usually shorter than those containing combination of all operators. In redescriptions with queries containing only conjunction operators, every attribute used in its queries must describe all instances from redescription support set. Thus, such redescriptions discover stronger associations between attributes than redescriptions with queries containing all operators. These reasons make us believe that applying CLUS-RM with restriction to use of conjunctions to ADNI data is the right choice which may reveal useful medical hypotheses that can be further developed by the domain experts. Described query language is similar to the one used in bi-directional association rules which can, for instance, be produced by the two-view data association discovery approach, discussed in the Introduction section. In general, using negation and disjunction operators in redescription construction can increase the diversity and accuracy of produced redescriptions, but it can also make them more difficult to understand for domain experts.

## CLUS-RM algorithm description

All experiments were performed with the CLUS-RM redescription mining algorithm [29, 30], presented in Algorithm 1. CLUS-RM uses PCTs [43] to find descriptions of groups of instances (i.e. subjects, as is the case in our medical study).

**Algorithm 1** The CLUS-RM algorithm

**Require:** First view ($W_1$), Second view ($W_2$), maxIter, Quality constraints $\mathcal{Q}$
**Ensure:** A set of redescriptions $\mathcal{R}$
1: **procedure** CLUS-RM
2:    $[W_1^{(0)}, W_2^{(0)}] \leftarrow$ createInitalData($W_1, W_2$)
3:    $[P_{W_1^{(0)}}, P_{W_2^{(0)}}] \leftarrow$ createInitialPCTs($W_1^{(0)}, W_2^{(0)}$)
4:    $[r_{W_1^{(0)}}, r_{W_2^{(0)}}] \leftarrow$ extractRulesFromPCT($P_{W_1^{(0)}}, P_{W_2^{(0)}}$)
5:    **for** Ind $\in \{1, \ldots,$ maxIter$\}$ **do**
6:       $[W_1^{(Ind)}, W_2^{(Ind)}] \leftarrow$ constructTargets($r_{W_1^{(Ind-1)}}, r_{W_2^{(Ind-1)}}$)
7:       $[P_{W_1^{(Ind)}}, P_{W_2^{(Ind)}}] \leftarrow$ createPCTs($W_1^{(Ind)}, W_2^{(Ind)}$)
8:       $[r_{W_1}^{(Ind)}, r_{W_2}^{(Ind)}] \leftarrow$ extractRulesFromPCT($P_{W_1^{(Ind)}}, P_{W_2^{(Ind)}}$)
9:       **for** ($R_{new} \in r_{W_1^{(Ind)}} \times_\mathcal{Q} r_{W_2^{(Ind-1)}} \cup r_{W_1^{(Ind-1)}} \times_\mathcal{Q} r_{W_2^{(Ind)}}$) **do**
10:          $\mathcal{R} \leftarrow$ addReplaceDiscard($R_{new}, \mathcal{R}$)
11:    $\mathcal{R} \leftarrow$ minimizeQueries($\mathcal{R}$)
12:    **return** $\mathcal{R}$

The presented algorithm pseudocode describes the CLUS-RM functionality in case only conjunction logical operators are used to create redescription queries. The extended version of the algorithm pseudocode for the case in which conjunction, negation and disjunction logical operators can be used in redescription query construction is described in [30] and supplementary document S18 File.

The algorithm consists of four main parts: 1) Initialization, 2) Query creation (divided in query construction 2.1 and query exploration 2.2), 3) Redescription creation and 4) Redescription set optimisation.

1) In the initialization phase (line 2 in Algorithm 1), the algorithm makes a copy of each instance from the original dataset and shuffles the attribute values for the copies. For each attribute, the algorithm selects a random instance from the dataset and copies its value for the selected attribute to the target copy (value of one instance from the original dataset can be copied multiple times). This procedure breaks correlations between attributes in the copied instances. Each instance from the original dataset is assigned a target value 1.0 and each artificially created instance a target value 0.0. It is possible to use the PCT algorithm to create initial clusters, from such dataset, by distinguishing between original instances and copies containing shuffled values (line 3 in Algorithm 1). The described procedure is repeated independently for both views contained in the dataset.

2.1) Each node in the obtained PCTs represents a cluster. These nodes are transformed to rules (line 4 in Algorithm 1) which are valid for the corresponding group of instances. More details about transforming PCTs to rules can be seen in [49].

2.2) The next step is to describe the same groups of instances, as those described by the produced rules, with the second attribute set (lines 6–8 in Algorithm 1). To do this, for each instance of the original dataset, the algorithm constructs a set of target variables containing equal number of targets as number of rules constructed using the first set of attributes (for more details see [30]). The instance has a target value 1 on position $j$ if it is described by the $j$-th rule from a set of rules constructed on the first set of attributes, otherwise the value is 0. Instances for which information is missing, making it impossible to determine the membership in support set of the query are also labelled with 0. We use the multi-target classification and regression capability of PCT to construct clusters on different views containing similar instances. The procedure is repeated by creating initial rules on the second view and describing similar sets of instances by using attributes from the first view.

3) Once the algorithm obtains rules for both views, it combines them by computing the Cartesian product of two rule-sets (line 9 in Algorithm 1). Each redescription is evaluated with various user predefined constraints (such as minimal redescription accuracy, minimal support, maximal *p*-value, contained in a set of redescription quality constraints $\mathcal{Q}$), to select candidates for redescription set optimization.

4) Each redescription satisfying all user-defined redescription quality constraints is a candidate for redescription set optimization (line 10 in Algorithm 1). Satisfactory redescriptions are added to the redescription set, in the order of creation, until the maximal number of redescriptions (user-defined parameter) is reached. When this number is reached, the algorithm computes the score difference, defined in [29, 30], between the new redescription and every redescription already contained in the redescription set based on redescription score. The score of some redescription $R \in \mathcal{R}$, based on its support set and a redescription set $\mathcal{R}$, is computed as:

$$redScoreInst(R) = \frac{\sum_{i \in supp(R)}(coverInst_{\mathcal{R} \setminus R}(i))}{\sum_{i \in D} coverInst_{\mathcal{R}}(i)}$$

where $coverInst_{\mathcal{R}}(i) = |\{R \in \mathcal{R}, i \in supp(R)\}|$ denotes the number of times, the instance *i* is described by redescriptions from the redescription set $\mathcal{R}$. The denominator of a score *redScoreInst(R)* can be also written as $\sum_{R \in \mathcal{R}}|supp(R)|$. Similarly, the redescription score:

$$redScoreAttr(R) = \frac{\sum_{a \in attr(R)}(coverAttr_{\mathcal{R} \setminus R}(a))}{\sum_{a \in attr(D)} coverAttr_{\mathcal{R}}(a)}$$

is based on attributes contained in redescription queries, where $coverAttr_{\mathcal{R}}(a) = |\{R \in \mathcal{R}, a \in attr(R)\}|$ denotes the number of times attribute *a* is used in queries of redescriptions contained in $\mathcal{R}$. The denominator of a score *redScoreAttr(R)* can be also written as $\sum_{R \in \mathcal{R}}|attr(R)|$.

The score of a newly created redescription $R_{new}$ is computed in the same way as the score for some $R \in \mathcal{R}$ but using frequencies for all redescriptions contained in the set $\mathcal{R}$ in the numerator of *redScore* and *redScoreAttr*.

The error score is computed as $errSc(R) = 1.0 - J(R)$ and the final redescription score is computed as:

$$sc(R) = \alpha_1 \cdot errSc(R) + \alpha_2 \cdot redScoreInst(R) + \alpha_3 \cdot redScoreAt(R)$$

where $\alpha_i \in [0, 1]$, $\sum_{k=1}^{3} \alpha_i = 1$. Lower total redescription score is favourable because it implies smaller error in redescription accuracy and smaller level of instance and attribute redundancy with respect to other redescriptions from the set $\mathcal{R}$. The user—defined weights $\alpha_k$ regulate importance of different scores which affect the properties of the resulting redescription set. In this work, we use $\alpha_k = \frac{1}{3}$. Redescription contained in the redescription set with the highest score difference with the newly created redescription is replaced thus improving the overall redescription set quality. At each redescription exchange all frequency scores are updated.

The minimization procedure introduced in [30] and performed in line 11 of Algorithm 1 is a heuristic procedure designed to reduce the size of redescription queries by removing redundant attributes (attributes that can be removed without changing redescription accuracy). It is performed individually on each redescription of the resulting redescription set.

**Constraint-based redescription mining.** In this work, we extended the CLUS-RM algorithm to a constraint-based redescription mining setting. The algorithm incorporates constraints in redescription creation and one additional score in the optimization function used for redescription set creation. Necessary CBRM extensions of the CLUS-RM algorithm, when conjunction, negation and disjunction operator can be used in redescription query construction are described in supplementary document S18 File.

We present the attribute level constraints useful for gaining knowledge as demonstrated in this work. Constraints involving instances can be introduced in the analogous fashion by using redescription support set ($supp(R)$) instead of attribute set ($attrs(R)$) in formulas (1), (2) and (3).

Constraint-based redescription mining, first defined in [31], allows placing constraints on attributes that must occur in redescription queries or instances that must be contained in redescription support set. The constraints are in the form $\mathcal{C} = \{C_1, C_2, \cdots, C_n\}$, where each constraint $C_i$ specifies a set of attributes that must occur in redescription queries or a set of instances that must be contained in redescription support set. In the original formulation, at least one constraint $C_i$ must be satisfied by a redescription (contain all attributes or instances specified in the set) to be presented to the user. We denote this original definition as strict constrained-based redescription mining and mostly use it in our study. In practice, various relaxed versions of constrained-based redescription mining might be useful. In the continuation, we specify one existing (strict) and two newly defined (soft and suggested) modes of constraint-based redescription mining (focusing only on attribute constraints):

1. Strict constraint-based redescription mining: there must exist at least one constraint $C_i \in \mathcal{C}$ such that all defined attributes occur in redescription queries.

2. Soft constraint-based redescription mining: there must exist at least one constraint $C_i \in \mathcal{C}$ such that a part of defined attributes occurs in redescription queries. Satisfying larger portion of constraints is favoured by the redescription evaluation score.

3. Suggested constraint-based redescription mining: defined constraints are used as suggestions that increase the overall redescription score depending on the number of satisfied constraints, however high quality redescriptions not satisfying any of these constraints can also enter redescription set if their total score is high enough.

Strict constraint-based redescription mining can be used when the expert already has a hypothesis (obtained through domain knowledge and extensive experimentation) and wants to explore the specified associations in more detail. Soft constraint-based redescription mining can be used when a set of attributes of interest has been determined (by applying the combination of domain knowledge and experimentation) but it is not clear which interactions from the set should be fully explored. Thus, further study of their interactions is needed to form, refine or confirm the expert hypothesis. Suggested constraint-based redescription mining can be used when the expert, knowing the research domain (having a priori knowledge about the problem), selects a set of attributes that are known or suspected to be (currently) more interesting for exploration, though at current stage there is no immediate focus on any particular hypothesis.

To allow constraint-based redescription mining, we extend the CLUS-RM algorithm by adding a new set of constraints $\mathcal{C}$ containing the user-defined attributes of special interest and a type of CBRM used (parameter $\mathcal{T}$). Line 9 of Algorithm 1 is changed to $R_{new} \in (r_{W_1}^{(ind)})_{\{\mathcal{C}, \mathcal{T}\}} \times_Q (r_{W_2}^{(ind)})_{\{\mathcal{C}, \mathcal{T}\}}$. Thus, redescriptions are created only by combining those queries that satisfy predefined constraints. For each redescription $R_{new}$, we apply query minimization procedure before using redescription set optimization (defined in line 10 of Algorithm 1).

If query minimization procedure removes any of the key constraint attributes, defined in set $\mathcal{C}$ of CBRM, the created redescription is discarded.

In addition, CLUS-RM is extended with a new score *scConstr*, which is used in suggested constraint-based redescription mining to increase the overall score of a redescription satisfying user-defined attribute constraints. The score is defined as:

$$scConstr(R) = \frac{1}{2} \cdot max\left\{\frac{|attrs(R) \cap C_i|}{|C_i|}, \ C_i \in \mathcal{C}\right\} + \frac{1}{2} \cdot \frac{|attrs(R) \cap (\cup_i C_i)|}{|attrs(R)|} \quad (1)$$

The first term in the score rewards redescriptions satisfying higher fraction of constraints from some set $C_i$. Due to the fact that more disjoint or partially overlapping constraint sets can be given and the fact that some redescriptions can satisfy parts of larger number of constraint sets $C_i$, we take the maximum score achieved among constraint sets as a quality of redescription—thus favouring compliance with larger number of constraints from a single constraint set. The second term favours redescriptions that, among the attributes contained in their queries, have larger fraction of attributes of interest to the user. Here, we reward satisfied constraints from any constraint set defined by the user.

The score used for soft constraint-based redescription mining is defined as:

$$redScoreSoft = \begin{cases} scConstr(R) & \text{if } \exists C_i \in \mathcal{C}, attrs(R) \cap C_i \neq \emptyset \\ -\infty & \text{otherwise} \end{cases} \quad (2)$$

Similarly, the score used for strict constraint-based redescription mining is defined as:

$$redScoreStrict = \begin{cases} 1 & \text{if } \exists C_i \in \mathcal{C}, attrs(R) \cap C_i = C_i \\ -\infty & \text{otherwise} \end{cases} \quad (3)$$

Higher scores denote higher level of agreement of redescriptions with the imposed constraints (redescriptions with higher score are thus preferable).

Finally, redescription score *sc(R)* is extended to:

$$\begin{aligned} sc(R) &= \alpha_1 \cdot errSc(R) + \alpha_2 \cdot redScoreInst(R) + \\ &\quad + \alpha_3 \cdot redScoreAt(R) + \alpha_4 \cdot (1 - redScoreConst(R)) \end{aligned}$$

where $\alpha_i \in [0, 1]$, $\sum_{k=1}^{4} \alpha_i = 1$ and *redScoreConst(R)* denotes any variant of the constraint-based score chosen by the user. Redescriptions with the score value of $\infty$ are not allowed to enter redescription set.

With the extension introduced above, the CLUS-RM is the only redescription mining algorithm capable of performing fully automated constraint-based redescription mining on categorical, numerical and data containing missing values with more than one attribute constraint.

## Experiments and results

In this section, we present the experimental setup and some selected results obtained through the analyses of the produced redescription sets.

### Experiments

Our main goal was to study clinical and biological attributes, and to find interesting relations among them. To retrieve maximum information from and to obtain deeper insight into the data, we divided redescriptions by the number of described subjects and used the diagnosis of the level of cognitive impairment to further assess the relevance and interestingness of the obtained redescriptions. For each dataset, we created four redescription sets containing redescriptions with different supports, describing [5, 10], [11, 39], [40, 99] and at least 100 subjects. The maximum support threshold was set to $\lceil \frac{|D_i|}{2} \rceil$ subjects contained in the dataset $D_i$, $i \in \{1, 2, 3\}$. We are interested in re-describing subsets of subjects with some level of cognitive impairment and using cognitively normal subjects as a control group. Studying different biological, clinical attributes and their interactions in the context of different levels of cognitive impairment is also of high interest. Higher homogeneity of described subjects increases the amount of information obtained about different changes in biological and clinical attributes occurring as a result of different level of cognitive impairment. Developing an approach with a combined properties of redescription mining and subgroup discovery may also be interesting in this setting, but is beyond the scope of this work. Each set contains 100 redescriptions with a minimal Jaccard Index of 0.2 and a maximal *p*-value of 0.01. Allowed support intervals, as well as other parameter limits were found through experimentation. Redescriptions contained up to 8 attributes per query.

The same support intervals were used to create redescriptions on each dataset. This allows making easier comparisons of redescriptions and statistics of attribute co-occurrence across different datasets. Distribution analysis of redescription quality measures, in the produced redescription sets, reveals potentially interesting datasets, attributes and support intervals.

Since PAPP-A showed interesting associations with cognitive impairment in the experiments described above, we performed constraint-based redescription mining with the same algorithmic parameters but focusing redescription search on redescriptions containing pregnancy associated plasma protein A (PAPP-A) in the redescription queries. We created one redescription set containing 100 redescriptions describing at least 100 subjects.

### Redescription accuracy and homogeneity analysis

We merged the four sets of redescriptions, of different supports, created on each dataset ($D_1$, $D_2$, $D_3$) and formed one larger redescription set (RS) per dataset, denoted $\mathcal{R}_1, \mathcal{R}_2, \mathcal{R}_3$ (see Fig 2). For the obtained redescriptions, contained in the corresponding redescription sets ($\mathcal{R}_1, \mathcal{R}_2, \mathcal{R}_3$), we analysed the homogeneity of the described subsets of subjects with respect to the degree of cognitive impairment (CN, SMC, EMCI, LMCI and AD) by computing the entropy of described subject's medical diagnosis (demonstated in Fig 2).

The entropy was computed for the support set of each redescription by using the package *entropy* developed for the programming language *R*. The package allows estimating Shannon's entropy ($H = -\sum_{i=0}^{N-1} p_i \cdot log_2(p_i)$) [50] of some finite set of probabilities obtained from the observed counts (occurrence frequencies of each level of cognitive impairment in the redescription support set). In this use-case, $N$ equals the number of different target classes occurring in the support set of a redescription. Probability $p_i$ is computed as $p_i = \frac{|target_i \cap supp(R)|}{|supp(R)|}$, where $target_{i, i \in \{0, ..., N-1\}}$ denotes a set of entities with target label *i*. Characteristics of redescription sets produced with different support intervals (1., 2., 3., 4. in Fig 2), can be seen on a plot showing entropy distributions (i in Fig 2) and distributions of redescriptions' Jaccard index (ii in Fig 2).

Due to the smaller diversity in target classes (containing no SMC subjects and only 1 EMCI subject), it was easier to distinguish between different groups of subjects on dataset 3 (which is illustrated in Fig 2) than on the other two datasets. On dataset 3, we obtained many clusters of various size, homogeneous with respect to medical diagnosis, which gives us confidence that we found attribute combinations and numerical intervals useful for the analysis and understanding of cognitive impairment connected to AD.

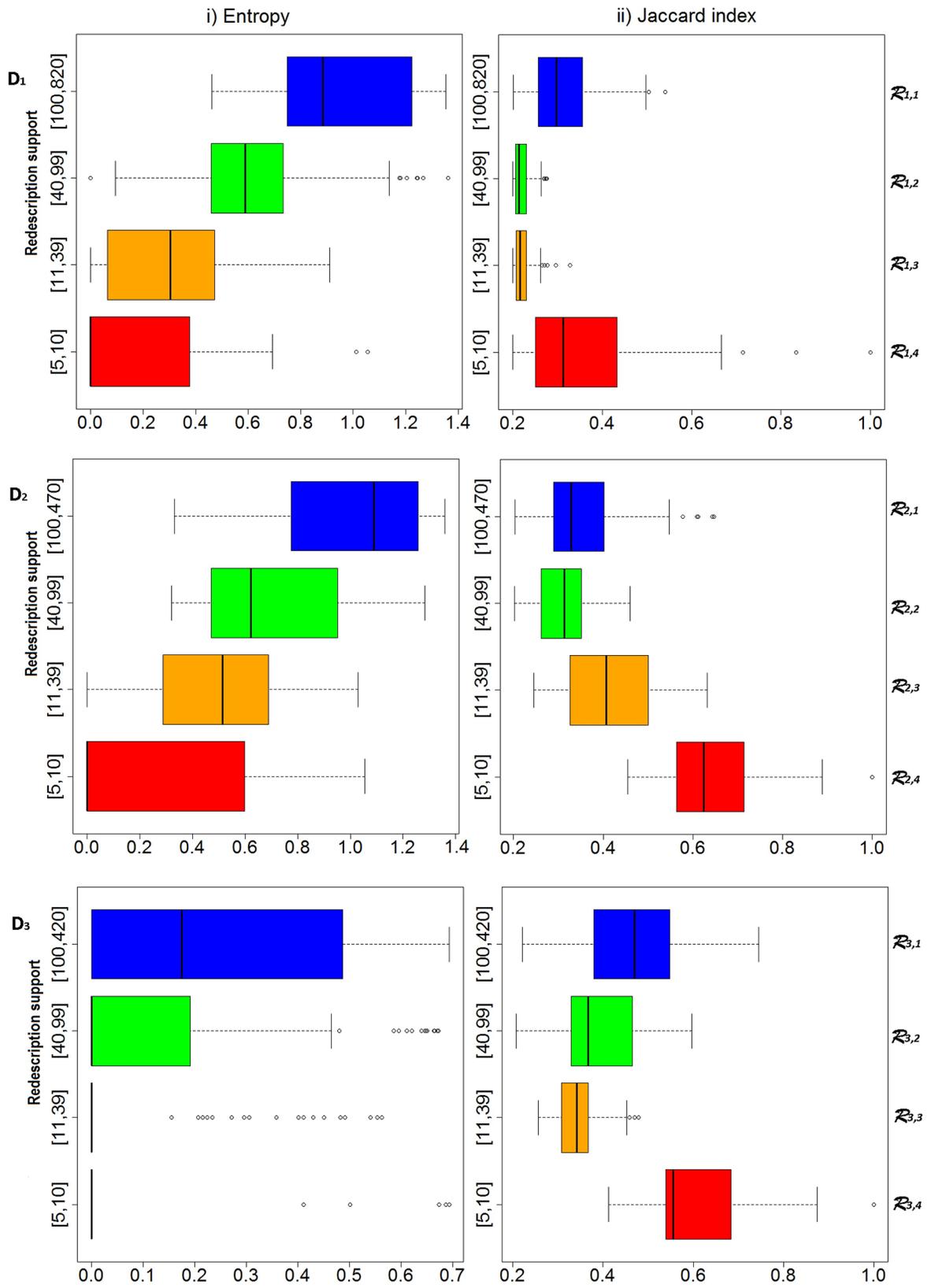

**Fig 2. Entropy (i) and Jaccard index (ii) value distributions for the redescription sets created on each dataset (first dataset—$D_1$ at the top, third dataset—$D_3$ at the bottom).** For a dataset $D_i$, $i \in \{1, 2, 3\}$, we create four redescription sets $\mathcal{R}_{i,1} - \mathcal{R}_{i,4}$ so that the number of described subjects in each redescription (from a particular redescription set) falls in the corresponding interval shown on the y-axis (boxplots representing distributions for each interval are coloured in different color). Each redescription set $\mathcal{R}_{i,j}$, $i \in \{1, 2, 3\}$, $j \in \{1, \cdots, 4\}$ contains 100 redescriptions.

https://doi.org/10.1371/journal.pone.0187364.g002

The entropy increases with the increase of the number of described subjects, while the Jaccard index shows stronger associations in redescriptions with support in the first ($|supp(R)| \geq 100$ in Fig 2) and the last interval ($|supp(R)| \in [5, 10]$ in Fig 2). Redescriptions describing the smallest number of subjects (the last interval) use larger number of attributes with very specific numerical intervals to isolate groups of subjects that are very homogeneous with respect to the medical diagnosis and describe many different groups of subjects suffering from severe cognitive impairment (LMCI, AD). In contrast, many accurate redescriptions (in the first interval) use larger numerical intervals, thus often describing subjects with various levels of cognitive impairment. Additional reason for higher accuracy in this interval compared to the middle two intervals is the detection of highly accurate redescriptions describing subgroups of CN subjects. Missing values in the data and potential noise, occurring from the errors in measurements and data processing, negatively affect the Jaccard index.

### Analyses based on examination of redescription sets

Redescription set analyses, which included: a) the examination and expert evaluation of individual redescriptions, b) the distribution analysis of level of dementia for the described subjects of these redescriptions, c) comparative analyses of attribute value distribution between different subsets of subjects (LMCI/AD vs CN or $supp(R)$ vs CN), allowed us to find useful information related to subjects with cognitive impairment.

From the clinical attributes, we noticed that ADAS, MOCA, Geriatric Depression Scale, Rey Auditory Verbal Learning Test (especially the percent forgetting score), and Mini-Mental State Exam (MMSE) occurred frequently in queries of obtained redescriptions that describe subjects suffering from various degrees of cognitive impairment. Nevertheless, there were instances where some CN subjects fell in the identified intervals of values for these attributes. Attributes connected to Clinical Dementia Rating distinguished well between CN subjects and those with different degrees of cognitive impairment. Redescriptions mostly contained the attributes CDMEMORY, CDGLOBAL and CDR-SB (clinical dementia rating sum of boxes). From the biological attributes, we often encountered attributes connected to brain volume, hippocampus, various blood and urinary tests (attributes HMT and RCT), intracranial volume (ICV), $^{18}$fluorodeoxyglucose—positron emission tomography (FDG-PET) and $^{18}$F-florbetapir (AV45). These attributes have been studied before by Gamberger et al. [3, 13]. We noticed that the biological attribute SPARE_AD (Spatial Pattern of Abnormalities for Recognition of Early AD) correlated with subject's diagnosis very well and occurred frequently in redescriptions constructed on dataset 3 that contains it. Also, the gene variant *APOE ε4* was present exclusively in redescriptions describing subjects diagnosed with LMCI and AD.

We report several attributes, discovered during our analyses, for which we detected variations in levels connected to AD or discovered interesting subgroups of patients with significantly different distribution of values for a given attribute compared to CN subjects. Difference in distribution is measured with three different statistical tests: a) Anderson-Darling (ADT) test [51, 52], Kolmogorov-Smirnov (KST) test [53, 54] and Mann-Whitney U (MWUT) test [55]. For Anderson-Darling we perform two-sided test and report simulated ($p_s$) and asymptotic ($p_a$) $p$-values, while for Kolmogorov-Smirnov and Mann-Whitney U test we report $p$-values, obtained by performing one-sided tests, and the observed direction of the shift of distribution. Alternative hypothesis (a), for one-tailed tests have two possible forms: a equals (=) less (l), or (a) = greater (g). Depending on the choice of statistical test, the alternative hypothesis have different meaning (explained in S17 File). Simulated $p$-value in ADT are obtained with default parameters (1000 simulations). Short motivation for the used statistical tests, providing references to implementations and meaning of the chosen alternative, for the used one-sided tests, is available in supplementary document S17 File. Tests of statistical significance of difference in distribution between one selected example group and a group of CN subjects for all mentioned attributes is displayed in Table 5. Information about attributes with statistically significant difference in distribution between AD/LMCI and CN subjects is reported in Table 6.

By observing redescriptions describing very homogeneous groups of subjects with high level of cognitive impairment (LMCI and AD), we discovered groups where testosterone levels (TSTSTRNT) were significantly decreased. Although some studies (e.g. Zhao et al. [56]) and meta-analyses showed no differences in plasma levels of testosterone between AD and matched controls (e.g. Xu et al. [57]), some studies, such as the one of Hogervorst et al. [58] and Lv et al. [59], found low free testosterone level to be an independent risk factor for AD. Plasma testosterone levels display circadian variation, peaking during sleep, and reaching a lowest level in the late afternoon, with a superimposed ultradian rhythm with pulses every 90 min reflecting the underlying rhythm of pulsatile luteinizing hormone (LH) secretion [60]. The increase in testosterone during sleep requires at least 3 hours of sleep with normal sleep architecture. However, since noradrenergic locus coeruleus and serotonergic dorsal raphe nucleus are among the first neurons affected by neurofibrillary tau pathology, their changes lead to the early and prominent deterioration of the sleep-wake cycle in AD (for a review, see Šimić et al. [61]), which may add to a reduction of testosterone levels with advancing age. Experimental data obtained in animal models of AD suggest that low levels of testosterone increase A$\beta$ and tau pathology through both androgen and estrogen pathways (testosterone is metabolized in the brain into androgen dihydrotestosterone, DHT, and 17$\beta$-estradiol, the E2 estrogen) [62, 63].

**Table 5. Attributes analysed in this section with corresponding example redescription containing this attribute.** For each selected attribute we present example redescription that describes subjects with statistically significant difference in attribute value distribution compared to a group of CN subjects.

| Attribute | D | R | $|E_{1,1}|$ | File | ADT | | KST | | MWUT | |
|---|---|---|---|---|---|---|---|---|---|---|
| | | | | | $p_a$ | $p_s$ | a | p | a | p |
| ANG2 | $D_3$ | $R_{45}$ | 46 | S14 | $4.1 \cdot 10^{-3}$ | 0 | l | $2.7 \cdot 10^{-6}$ | g | $4.7 \cdot 10^{-6}$ |
| APOAII | $D_3$ | $R_{37}$ | 55 | S14 | $7.3 \cdot 10^{-15}$ | 0 | g | $1.7 \cdot 10^{-11}$ | l | $4.2 \cdot 10^{-13}$ |
| BNP | $D_3$ | $R_{96}$ | 48 | S14 | $5.7 \cdot 10^{-3}$ | 0 | l | 0.02 | g | 0.15 |
| CNTF | $D_3$ | $R_{56}$ | 33 | S13 | 0.03 | 0.03 | l | 0.02 | g | 0.02 |
| TSTSTRNT | $D_3$ | $R_{85}$ | 366 | S15 | $5 \cdot 10^{-6}$ | 0 | g | 0.002 | l | 0.05 |
| INSULIN | $D_3$ | $R_{90}$ | 5 | S12 | 0.01 | 0.01 | l | 0.06 | g | 0.01 |
| LEPTIN | $D_3$ | $R_{72}$ | 24 | S13 | $9.4 \cdot 10^{-6}$ | 0 | g | $5.1 \cdot 10^{-6}$ | l | $7.4 \cdot 10^{-6}$ |
| MCRPHMIF | $D_3$ | $R_{31}$ | 6 | S12 | $9 \cdot 10^{-5}$ | 0 | l | $4.2 \cdot 10^{-4}$ | g | $2.3 \cdot 10^{-4}$ |
| PAPP-A | $D_3$ | $R_{39}$ | 327 | S16 | $3 \cdot 10^{-6}$ | 0.0 | l | $4.8 \cdot 10^{-4}$ | g | $1.8 \cdot 10^{-5}$ |
| PPP | $D_3$ | $R_{43}$ | 8 | S12 | $8.8 \cdot 10^{-3}$ | 0.13 | l | 0.02 | g | 0.008 |
| SPARE_AD | $D_3$ | $R_{37}$ | 155 | S15 | $1.2 \cdot 10^{-28}$ | 0.0 | l | $2.2 \cdot 10^{-16}$ | g | $2.2 \cdot 10^{-16}$ |

https://doi.org/10.1371/journal.pone.0187364.t005

**Table 6. Analysed attributes with statistically significant difference in value distribution between groups of LMCI or AD patients and CN subjects.**

| Attribute | D | Type | ADT | | KST | | MWUT | |
|---|---|---|---|---|---|---|---|---|
| | | | $p_a$ | $p_s$ | a | p | a | p |
| APOAII | $D_3$ | LMCI vs CN | $4.6 \cdot 10^{-11}$ | 0 | g | $1.1 \cdot 10^{-9}$ | l | $3.1 \cdot 10^{-10}$ |
| | $D_3$ | AD vs CN | $3.3 \cdot 10^{-7}$ | 0 | g | $8.4 \cdot 10^{-5}$ | l | $3.3 \cdot 10^{-7}$ |
| APOB | $D_3$ | AD vs CN | 0.03 | 0.04 | l | 0.03 | g | 0.02 |
| ANG2 | $D_3$ | LMCI vs CN | $2.6 \cdot 10^{-4}$ | 0 | l | $4.8 \cdot 10^{-3}$ | g | $1.5 \cdot 10^{-4}$ |
| BNP | $D_3$ | LMCI vs CN | $9.2 \cdot 10^{-8}$ | 0 | l | $1.8 \cdot 10^{-5}$ | g | $1.2 \cdot 10^{-6}$ |
| | $D_3$ | AD vs CN | $6 \cdot 10^{-7}$ | 0 | l | $1.3 \cdot 10^{-5}$ | g | $1.2 \cdot 10^{-6}$ |
| FASL | $D_3$ | LMCI vs CN | $3 \cdot 10^{-5}$ | 0 | g | 0.001 | l | $2 \cdot 10^{-5}$ |
| LEPTIN | $D_3$ | LMCI vs CN | $1.2 \cdot 10^{-3}$ | 0 | g | $6 \cdot 10^{-3}$ | l | $4.7 \cdot 10^{-4}$ |
| | $D_3$ | AD vs CN | 0.05 | 0.05 | g | 0.08 | l | 0.02 |
| PAPP-A | $D_3$ | LMCI vs CN | $7.2 \cdot 10^{-4}$ | 0.001 | l | $1.3 \cdot 10^{-3}$ | g | $3.4 \cdot 10^{-4}$ |
| | $D_3$ | AD vs CN | $6.1 \cdot 10^{-6}$ | 0 | g | $1.1 \cdot 10^{-4}$ | l | $8.3 \cdot 10^{-5}$ |
| PPP | $D_3$ | LMCI vs CN | $6.2 \cdot 10^{-3}$ | 0.005 | l | 0.009 | g | 0.003 |
| | $D_3$ | AD vs CN | $2.5 \cdot 10^{-3}$ | 0.001 | l | 0.007 | g | $1.5 \cdot 10^{-3}$ |

https://doi.org/10.1371/journal.pone.0187364.t006

Unlike previous scarce data and negative correlation [64], we also found increased levels of ciliary neurotrophic factor (CNTF) in plasma in several redescriptions describing subjects with high level of cognitive impairment, together with decreased levels of leptin. The difference in distribution of leptin level between groups of AD/LMCI patients and CN subjects is significantly different (lower for AD and LMCI patients). This is in agreement with the results of Marwarha and Ghribi [65], showing that lower leptin levels detected in AD subjects can be a possible target for developing supplementation therapies for reducing the progression of AD. Some groups of subjects (such as $R_{45}$ from S14 File) had significantly increased levels of plasma angiopoietin-2 (ANG2). This is in agreement with research by Thirumangalakudi et al. [66] and research by Grammas et al. [67], that revealed elevated expression of angiopietin-2 in the brains of AD subjects and the transgenic AD mice, respectively.

Increased levels of plasma brain natriuretic peptide (BNP) were found in several redescriptions containing subjects with severe cognitive impairment. Previous research [68] suggested that this peptide has more significant association with vascular dementia than with AD. This could suggest either that this group of subjects, described by redescriptions containing BNP attribute, suffered from both types of dementia (mixed dementia), or that these cases do not suffer from AD but indeed suffer from vascular dementia. Distributions of level of BNP are significantly different, in dataset $D_3$, between groups of LMCI/AD and CN subjects.

Finally, we also found alteration in plasma levels of several other attributes, whose relationship with AD has already been shown in the literature. These include increase in serum apolipoprotein B (APOB) [69], pancreatic polypeptide (PPP) [70, 71] and for very small groups, the increase of plasma insulin [72] and the CSF macrophage migration inhibitory factor (MCRPHMIF) [73] in AD brain. Fas (CD95) ligand (FASL) levels are found to be significantly decreased in LMCI patients compared to AD and CN subjects in our dataset. Levels are also lower in AD patients than in CN subjects but the difference is not statistically significant. Although one study suggests the upregulation of FASL in AD brain [74], the levels and variations seem to heavily depend on the part of the brain. For instance, FASL levels are found to be significantly decreased in hippocampus [75] in patients suffering from AD. Several groups of LMCI/AD patients with significantly lower levels of APOAII compared to the CN subjects were detected (which corresponds to research performed in [76, 77]). The difference in value

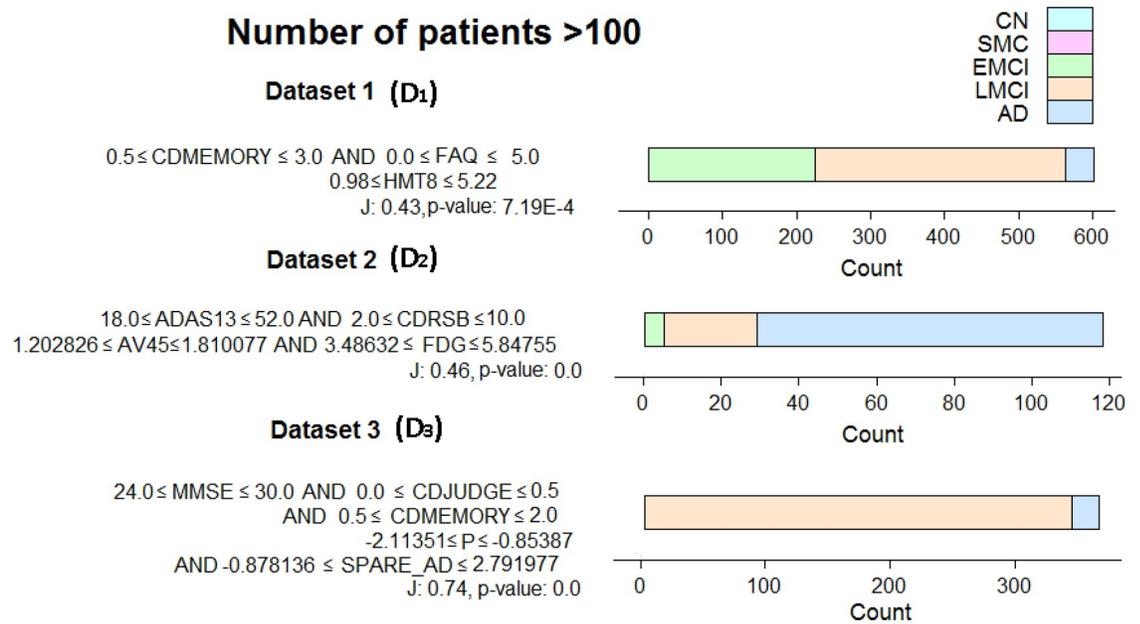

**Fig 3. Example redescriptions (one for each dataset), each describing at least 100 subjects.** All subjects described are diagnosed with EMCI, LMCI or AD. Attribute explanations can be seen in Tables 2 and 3 (P denotes PAPP-A and FDG denotes FDG-PET).

https://doi.org/10.1371/journal.pone.0187364.g003

distribution in dataset $D_3$ is significant between groups of LMCI/AD patients and CN subjects. Alterations in the levels of the PAPP-A attribute between CN subjects and LMCI/AD patients are very interesting (see Tables 5 and 6). The PAPP-A levels rise in LMCI subjects than drop significantly in AD subjects. This very property has been already detected in [78].

For each redescription set, we extracted one interesting, statistically significant redescription, and displayed its queries, along with the diagnosis distribution of the subjects described by this redescription (as shown in Fig 3).

The three redescriptions (as shown in Fig 3 from top to bottom) describe 602, 118 and 365 subjects, respectively with different proportion of EMCI, LMCI and AD diagnosis. They are statistically significant and describe 46%, 20% and 62% of all subjects with some level of cognitive impairment contained in the corresponding dataset. Their queries mostly contain well known attributes listed in Table 2 and in S1–S3 Files. The clinical attributes contained are memory score (CDMEMORY), Clinical Dementia Rating Scale Sum of Boxes (CDRSB), judgement and problem solving score (CDJUDGE), Alzheimer's Disease Assessment Scale (ADAS), Mini-Mental State Exam (MMSE). The biological attributes used contain neutrophils (HMT8), $^{18}$F-florbetapir (AV45), $^{18}$fluorodeoxyglucose—positron emission tomography (FDG-PET), Spatial Pattern of Abnormalities for Recognition of Early Alzheimer's disease (SPARE_AD) and Pregnancy-Associated Protein-A (PAPP-A) measurements.

## Pairwise attribute association analysis based on co-occurrences

In this section, we present results of attribute association analyses based on attribute co-occurrences in queries of redescriptions contained in our redescription sets. To obtain these associations, we studied the attribute co-occurrence frequencies in redescriptions contained in redescription sets $\mathcal{R}_1$, $\mathcal{R}_2$ and $\mathcal{R}_3$. We focused on redescriptions describing subjects with some level of cognitive impairment. Co-occurrence frequencies were computed separately for pairs of attributes contained in views bio-bio, clin-clin and bio-clin, where *bio* denotes the view containing biological and *clin* denotes the view containing clinical attributes. Finally, we merged all redescriptions computed on all three datasets to obtain global information about pairwise attribute associations (set $\mathcal{R}_1 \cup \mathcal{R}_2 \cup \mathcal{R}_3$). We do this for bio-bio, clin-clin and bio-clin combinations of views. Besides the associations, we also computed the pairwise attribute correlations, by using values of all subjects in the corresponding dataset for the selected pair of attributes, and the statistical significance of these correlations. For each attribute we performed the Kolmogorov-Smirnov test to assess if its values, for subjects contained in the dataset, follow normal distribution. If we obtained *p*-values smaller than 0.05 for both attributes in the pair, we computed Pearson correlation coefficient [79], otherwise we computed the Spearman's correlation coefficient [80] and the appropriate *p*-value of the corresponding significance test. Spearman's test was also used to compute correlations involving attributes with ordinal values.

A short list of top 5 pairwise associations (by co-occurrence) between attributes contained in the analysed datasets is provided in Tables 7, 8 and 9.

Table 7 shows high association between FDG-PET and the volume of the hippocampus, the entorhinal cortex, as well as an attribute related to the volume of the lateral ventricles. High association was also found between intracranial volume and creatine kinase levels (CKMB). This enzyme is present in greatest amounts in skeletal muscle, myocardium, and brain. The FDG-PET attribute often occurred in the same descriptive rules as the attribute measuring the level of vitamin B12 (BAT126). Administering of vitamin B12 is known to have beneficial effects on cognition when there is insufficient level of B12 in the organism [81, 82]. The incidence of AD increases with age and in fact, older adults often show deficiency of vitamin B12,

**Table 7. The top five associations between pairs of biological attributes as measured by their co-occurrence in redescription queries.** Attribute correlations for a redescription set $\mathcal{R}_i$ are computed on dataset $D_i$. *P* denotes the Pearson correlation coefficient and *S* denotes the Spearman's correlation coefficient. $\mathcal{R}_u = \mathcal{R}_1 \cup \mathcal{R}_2 \cup \mathcal{R}_3$. Correlations for attribute pairs from the redescription set $\mathcal{R}_u$ are computed on the largest dataset containing both attributes.

| **Pairwise associations and correlations between biological attributes** | | | | | |
|---|---|---|---|---|---|
| **RS** | **Attribute pair** | **Co-occurrence** | **Test** | **Correlation** | ***p*-value** |
| $\mathcal{R}_1$ | Hippocampus, FDG-PET | 111 | P | 0.42 | $<2.2 \cdot 10^{-16}$ |
| | FDG-PET, Entorhinal | 106 | P | 0.35 | $<2.2 \cdot 10^{-16}$ |
| | FDG-PET, Ventricles | 52 | S | −0.39 | $<2.2 \cdot 10^{-16}$ |
| | FDG-PET, ICV | 46 | S | −0.39 | $<2.2 \cdot 10^{-16}$ |
| | FDG-PET, AV45 | 42 | S | −0.37 | $<2.2 \cdot 10^{-16}$ |
| $\mathcal{R}_2$ | FDG.PET, Hippocampus | 86 | P | 0.4 | $<2.2 \cdot 10^{-16}$ |
| | FDG-PET, Entorhinal | 76 | P | 0.31 | $<2.2 \cdot 10^{-16}$ |
| | FDG-PET, AV45 | 52 | S | −0.37 | $<2.2 \cdot 10^{-16}$ |
| | FDG-PET, RCT14 | 45 | S | 0.124 | 0.0003 |
| | FDG-PET, BAT126 | 31 | S | −0.007 | 0.42 |
| $\mathcal{R}_3$ | SPARE_AD, PAPP-A | 66 | S | −0.05 | 0.1 |
| | SPARE_AD, Entorhinal | 39 | S | −0.51 | $<2.2 \cdot 10^{-16}$ |
| | PLMNRARC, PAPP-A | 18 | S | −0.05 | 0.14 |
| | SPARE_AD, TNC | 17 | S | 0.09 | 0.14 |
| | PAPP-A, Entorhinal | 15 | S | 0.08 | 0.039 |
| $\mathcal{R}_u$ | Hippocampus, FDG-PET | 197 | P | 0.42 | $<2.2 \cdot 10^{-16}$ |
| | FDG-PET, Entorhinal | 182 | P | 0.35 | $<2.2 \cdot 10^{-16}$ |
| | FDG-PET, AV45 | 94 | S | −0.37 | $<2.2 \cdot 10^{-16}$ |
| | FDG-PET, Ventricles | 68 | S | −0.39 | $<2.2 \cdot 10^{-16}$ |
| | FDG-PET, ICV | 67 | S | −0.39 | $<2.2 \cdot 10^{-16}$ |

https://doi.org/10.1371/journal.pone.0187364.t007

**Table 8. The top five associations between pairs of clinical attributes as measured by their co-occurrence in redescription queries.** Attribute correlations for a redescription set $\mathcal{R}_i$ are computed on dataset $D_i$. $P$ denotes the Pearson correlation coefficient and $S$ denotes the Spearman's correlation coefficient. $\mathcal{R}_u = \mathcal{R}_1 \cup \mathcal{R}_2 \cup \mathcal{R}_3$. Correlations for attribute pairs from the redescription set $\mathcal{R}_u$ are computed on the largest dataset containing both attributes.

| Pairwise associations and correlations between clinical attributes | | | | | |
|---|---|---|---|---|---|
| RS | Attribute pair | Co-occurrence | Test | Correlation | *p*-value |
| $\mathcal{R}_1$ | ADAS13, RAVLT | 52 | S | −0.8 | $<2.2 \cdot 10^{-16}$ |
| | ADAS13, Q13SCORE | 49 | S | 0.5 | $<2.2 \cdot 10^{-16}$ |
| | ADAS13, CDMEMORY | 48 | S | 0.5 | $<2.2 \cdot 10^{-16}$ |
| | ADAS13, FAQ | 45 | S | 0.67 | $<2.2 \cdot 10^{-16}$ |
| | RAVLT, CDMEMORY | 43 | S | −0.63 | $<2.2 \cdot 10^{-16}$ |
| $\mathcal{R}_2$ | MOCA, ADAS13 | 60 | S | −0.72 | $<2.2 \cdot 10^{-16}$ |
| | MOCA, EcogPtPlan | 30 | S | −0.28 | $<2.2 \cdot 10^{-16}$ |
| | MOCA, CDMEMORY | 27 | S | −0.58 | $<2.2 \cdot 10^{-16}$ |
| | ADAS13, MMSE | 24 | S | −0.64 | $<2.2 \cdot 10^{-16}$ |
| | ADAS13, CDRSB | 23 | S | 0.66 | $<2.2 \cdot 10^{-16}$ |
| $\mathcal{R}_3$ | ADAS13, CDMEMORY | 60 | S | 0.76 | $<2.2 \cdot 10^{-16}$ |
| | MMSE, CDMEMORY | 42 | S | −0.73 | $<2.2 \cdot 10^{-16}$ |
| | ADAS13, CDRSB | 34 | S | 0.76 | $<2.2 \cdot 10^{-16}$ |
| | ADAS13, MMSE | 30 | S | −0.71 | $<2.2 \cdot 10^{-16}$ |
| | FAQ, ADAS13 | 29 | S | 0.7 | $<2.2 \cdot 10^{-16}$ |
| $\mathcal{R}_u$ | ADAS13, CDMEMORY | 122 | S | 0.5 | $<2.2 \cdot 10^{-16}$ |
| | ADAS13, FAQ | 82 | S | 0.67 | $<2.2 \cdot 10^{-16}$ |
| | ADAS13, CDRSB | 79 | S | 0.72 | $<2.2 \cdot 10^{-16}$ |
| | ADAS13, RAVLT | 77 | S | −0.8 | $<2.2 \cdot 10^{-16}$ |
| | ADAS13, MMSE | 77 | S | −0.69 | $<2.2 \cdot 10^{-16}$ |

https://doi.org/10.1371/journal.pone.0187364.t008

mainly due to the impaired vitamin B12 uptake in the gastrointestinal tract [83]. AD patients also have increased homocysteine levels in the blood. Since homocysteine is directly associated with brain atrophy, it is possible that vitamin B12 supplementation (that reduces homocysteine levels) can actually slow the progression of brain atrophy [81]. However, since meta-analyses failed to prove [84, 85] the connection of vitamin B12 supplementation with homocysteine levels and improved cognition, further studies should be conducted to resolve this issue. The correlation between FDG-PET and B12 values in our dataset was not statistically significant, though it may be more pronounced on a subset of subjects (for instance those above a certain age). It has been reported [86] that diagnosis based on FDG-PET can lead to false diagnosis of AD, where subjects can be cognitively normal or have cognitive impairment due to a reversible cause.

The clinical attributes ADAS, MOCA, MMSE, CDR, FAQ and RAVLT co-occurred frequently. Interestingly, the question number 13 (number of targets hit) from the ADAS test occurred very frequently in redescription queries. In this task, the participants are required to cross-out specific digits from a long list of digits. High frequency co-occurrences and corresponding correlations for all aforementioned attributes can be seen in Table 8.

There was also a strong association of the ADAS, CDR and MOCA clinical attributes with FDG-PET and SPARE_AD, the volume of the entorhinal cortex and the hippocampus, and other biological attributes (see Table 9). Correlations between these attributes were statistically significant. One of the most interesting associations is that between CDRSB and PAPP-A which is used in screening tests for Down syndrome. CDRSB and PAPP-A negatively correlated (−0.15) and the correlation was statistically significant at the significance level of 0.01.

PLOS ONE

Relating clinical and biological characteristics of cognitively impaired and AD patients**Table 9. The top five associations between pairs of attributes consisting of a clinical and a biological attribute.** The association is measured as their co-occurrence in redescription queries. Attribute correlations for a redescription set $\mathcal{R}_i$ are computed on dataset $D_i$. $P$ denotes the Pearson correlation coefficient and $S$ denotes the Spearman's correlation coefficient. $\mathcal{R}_u = \mathcal{R}_1 \cup \mathcal{R}_2 \cup \mathcal{R}_3$. Correlations for attribute pairs from the redescription set $\mathcal{R}_u$ are computed on the largest dataset containing both attributes.

| | Pairwise associations and correlations between a biological and a clinical attribute | | | | |
|---|---|---|---|---|---|
| RS | Attribute pair | Co-occurrence | Test | Correlation | *p*-value |
| $\mathcal{R}_1$ | ADAS13, FDG | 197 | S | −0.58 | $<2.2 \cdot 10^{-16}$ |
| | ADAS13, Entorhinal | 99 | S | −0.49 | $<2.2 \cdot 10^{-16}$ |
| | ADAS13, Hippocampus | 96 | S | −0.54 | $<2.2 \cdot 10^{-16}$ |
| | ADAS11, FDG | 79 | S | −0.55 | $<2.2 \cdot 10^{-16}$ |
| | CDMEMORY, FDG | 69 | S | −0.49 | $<2.2 \cdot 10^{-16}$ |
| $\mathcal{R}_2$ | ADAS13, FDG | 142 | S | −0.56 | $<2.2 \cdot 10^{-16}$ |
| | MOCA, FDG | 124 | S | 0.49 | $<2.2 \cdot 10^{-16}$ |
| | ADAS13, Entorhinal | 52 | S | −0.38 | $<2.2 \cdot 10^{-16}$ |
| | MOCA, Hippocampus | 44 | S | 0.45 | $<2.2 \cdot 10^{-16}$ |
| | RAVLT, FDG | 42 | S | 0.48 | $<2.2 \cdot 10^{-16}$ |
| $\mathcal{R}_3$ | ADAS13, SPARE_AD | 131 | S | 0.68 | $<2.2 \cdot 10^{-16}$ |
| | CDMEMORY, SPARE_AD | 110 | S | 0.72 | $<2.2 \cdot 10^{-16}$ |
| | CDRSB, SPARE_AD | 58 | S | 0.7 | $<2.2 \cdot 10^{-16}$ |
| | MMSE, SPARE_AD | 51 | S | −0.62 | $<2.2 \cdot 10^{-16}$ |
| | CDRSB, PAPP-A | 43 | S | −0.15 | 0.0002 |
| $\mathcal{R}_u$ | ADAS13, FDG | 339 | S | −0.58 | $<2.2 \cdot 10^{-16}$ |
| | ADAS13, Entorhinal | 171 | S | −0.49 | $<2.2 \cdot 10^{-16}$ |
| | ADAS13, Hippocampus | 136 | S | −0.54 | $<2.2 \cdot 10^{-16}$ |
| | ADAS13, SPARE_AD | 131 | S | 0.68 | $<2.2 \cdot 10^{-16}$ |
| | MOCA, FDG | 124 | S | 0.49 | $<2.2 \cdot 10^{-16}$ |

https://doi.org/10.1371/journal.pone.0187364.t009

**Associations with PAPP-A.** Motivated by the statistically significant association between PAPP-A and CDRSB, we used constraint-based redescription mining to create a new redescription set (on dataset $D_3$) by focusing only on redescriptions containing PAPP-A as one of the attributes in the redescription queries (corresponding redescription set is presented in supplementary document S16 File). The associations from this redescription set, containing 100 redescriptions, are presented in Table 10. Support sets of all constructed redescriptions contained both male and female subjects with diagnosis LMCI and AD.

The associations presented in Table 10 show that PAPP-A occurs frequently in redescription queries together with the clinical tests CDMEMORY, CDRSB, MMSE and ADAS13. Correlations between PAPP-A and all these attributes were statistically significant at the significance level of 0.01. Interestingly, SPARE_AD and PAPP-A occurred in every redescription from the redescription set obtained with constraint-based redescription mining. As noted earlier, the correlation between these two attributes was not statistically significant when measured for all subjects in the dataset. However, the correlation (Spearman's $\rho = -0.096$) was statistically significant (with $p = 0.026$) when measured for subjects with AD and LMCI at the significance level of 0.05. The fact that every redescription in the set obtained with constraint-based redescription mining described exclusively subjects with AD and LMCI possibly explains the high frequency of association between those attributes and necessitates further exploration of the role of PAPP-A in AD and LMCI. Additionally, we found an interesting association between PAPP-A and two other biological attributes: the volume of the entorhinal

PLOS ONE | https://doi.org/10.1371/journal.pone.0187364  October 31, 2017  25 / 35

Table 10. The top four associations of PAPP-A with other attributes based on attribute pair occurrences in redescription queries obtained by using constraint-based redescription mining on dataset $D_3$. *S* denotes Spearman's correlation coefficient. The produced redescription set contains 100 different redescriptions.

| Associations of PAPP-A with biological attributes | | | | |
|---|---|---|---|---|
| **Attribute pair** | **Co-occurrence** | **Test** | **Correlation** | ***p*-value** |
| SPARE_AD, PAPP-A | 100 | S | −0.05 | 0.1 |
| Fusiform, PAPP-A | 21 | S | 0.11 | 0.01 |
| Entorhinal, PAPP-A | 20 | S | 0.08 | 0.039 |
| Hippocampus, PAPP-A | 13 | S | 0.01 | 0.4 |
| Associations of PAPP-A with clinical attributes | | | | |
| **Attribute pair** | **Co-occurrence** | **Test** | **Correlation** | ***p*-value** |
| CDMEMORY, PAPP-A | 85 | S | −0.11 | 0.0034 |
| CDRSB, PAPP-A | 51 | S | −0.15 | 0.00019 |
| MMSE, PAPP-A | 49 | S | 0.13 | 0.00088 |
| ADAS13, PAPP-A | 42 | S | −0.11 | 0.0056 |

https://doi.org/10.1371/journal.pone.0187364.t010

cortex and the volume of the fusiform gyrus (Fusiform). Correlations between PAPP-A and these biological attributes were statistically significant at the significance level of 0.05.

## Discussion

The redescription mining approach to segmenting high-dimensional datasets offers several advantages over classical clustering, subgroup discovery and association mining, such as the capability to generate relevant equivalence associations among combinations of attributes. We performed redescription mining experiments on three different datasets, created by extracting different sets of attributes from the ADNI database, and measured the redescription accuracy and the level of homogeneity (in terms of level of cognitive impairment) of the subjects described by each redescription. Basically, the main aim of our study has been to differentiate between cognitively normal subjects and those with some level of cognitive impairment, using clinical and biological attributes potentially related to AD. Our experiments over the constructed datasets were deliberately split into different support ranges in terms of subjects described with redescriptions to allow extracting general and specific, relevant AD-related information.

In this study, we found a number of surprisingly large and homogeneous groups and many smaller, more specific subgroups of subjects that are described with informative redescriptions, in a large extent confirming findings of previous works, corroborating some previously debatable findings or providing additional information about various attributes. After obtaining interesting associations with PAPP-A, we used the introduced extensions to the CLUS-RM algorithm to perform constraint-based redescription mining, allowing us to further explore associations of various attributes with PAPP-A. CLUS-RM is extended to perform fully automated constraint-based redescription mining on data containing either numerical, categorical attributes or missing values. In addition, it is equipped with soft and suggested CBRM capability, introduced in this work.

The clinical attribute CDR (CDMEMORY, CDGLOBAL and CDR-SB) was shown to be a very good attribute for differentiating CN subjects and subjects with some level of cognitive impairment. The gene variant *APOE ε4* was associated with subjects with high level of cognitive impairment (LMCI and AD), whereas the biological attribute SPARE_AD was highly correlated with the subject's diagnosis.

Additionally, high association of ADAS, CDR, and MOCA clinical attributes with FDG-PET, SPARE_AD, and the volume of the entorhinal cortex and hippocampus were shown. When describing homogeneous groups of subjects with high level of cognitive impairment (LMCI and AD), the decrease of testosterone plasma levels, CNTF plasma levels and increase of BNP plasma levels were observed. Likewise, changes in other biological attributes previously reported as being altered in AD, such as increase in levels of serum apolipoprotein B, pancreatic polypeptide, plasma insulin and Fas (CD95) were found.

Finally, probably the most important finding of this study was the detection of altered levels of those biological attributes, for subjects with cognitive impairment, that could have potential as therapeutic targets in AD, namely decreased leptin and increased angiopoietin-2 plasma levels. Decreasing leptin levels have been suggested to alleviate AD-related cellular changes in rabbit organotypic slices [87] and in human neuroblastoma cell culture [88, 89], suggesting that lowered leptin levels detected in AD subjects can be a possible target for developing supplementation therapies for reducing the progression of AD. The finding of increased angiopoietin-2 plasma levels in AD patients is in accordance with the study of Thirumangalakudi et al. [66], who showed that angiopoietin-2 is expressed by AD, but not control-derived microvessels, supporting the idea of targeting the angiogenic changes in the microcirculation of the AD brain as a potential therapeutic approach in AD [67]. Altogether, analysing redescriptions from all three different datasets allowed finding many different associations. Some of these associations, such as SPARE_AD and PAPP-A are novel and require more in depth analysis with the supervision of domain experts. The correlation between SPARE_AD and PAPP-A was not statistically significant when computed for all subjects contained in the dataset $\mathcal{R}_3$, but it was statistically significant when computed only for subjects with AD and LMCI at the significance level of 0.05. PAPP-A showed significant correlation with the volume of the Fusiform gyrus and the volume of the Entorhinal cortex—both already known as being associated with AD [90, 91]. Further, PAPP-A had statistically significant correlation to the most widely used clinical cognitive tests: ADAS, Mini-Mental State Examination and Clinical Dementia Rating Sum of Boxes.

It has been shown [92] by measuring the reference intervals of PAPP-A (in 52 healthy males and 74 healthy, non-pregnant women) that the reference intervals are <22.9 ng/mL for men and <33.6 ng/mL for non-pregnant women. PAPP-A levels of smokers were lower than that of non-smokers and there is a positive correlation between serum PAPP-A levels and subjects' age. The measured median value of PAPP-A in males 6.85 with the range [undetectable, 24, 40] ng/mL were significantly higher than the median of female subjects 3.4 with the measured range [undetectable, 36, 7] ng/ml. For both males and females, non-smokers had higher levels of PAPP-A than smokers. For males, the difference was statistically significant and for females, it was not. PAPP-A levels in pre-menopause women were lower than in the post-menopause women, however the difference was not statistically significant. In male subjects, the study found a significant correlation between subjects' age and the level of PAPP-A, however in female subjects this correlation was not statistically significant.

Our search (PubMed search on 3 March 2016.) by using the keywords *pappalysin-1/Pregnancy-associated plasma protein-A (PAPP-A)* and *Alzheimer's disease* revealed only one publication [93] that associates PAPP-A with depressive symptoms.

Results by Llano et al. [78] show that PAPP-A is among the most significant descriptors in plasma proteomic data for distinguishing between CN, MCI and AD patients by different supervised machine learning algorithms. We discovered associations between PAPP-A and cognitive status (LMCI, AD). These results demonstrate the importance of further study of PAPP-A as potential marker for early detection of AD.

PLOS ONE | Relating clinical and biological characteristics of cognitively impaired and AD patientsDistribution analysis of PAPP-A values based on our data and those of Llano et al. [78] show that PAPP-A levels are increased in MCI and LMCI patients but are significantly decreased in subjects diagnosed with AD. Decrease in PAPP-A levels from LMCI to AD patients on our data is more pronounced in female than in male patients. The possible link between PAPP-A and AD related genes (*ABCA1*, *ABCG1*) discovered in Hu et al. [94] is explained by Tang et al. [95]. This publication discusses the role of PAPP-A in pathogenesis of atherosclerosis through its inhibition of liver X receptors $\alpha$ (LXR$\alpha$) through the insulin-like growth factor (IGF)-I-mediated signalling pathway, and negative regulation of expression of *ABCA1* and *ABCG1* genes—all significantly associated with AD [94]. Although LXR are best known as the key regulators of cholesterol metabolism and transport, LXR signaling has also been shown to have significant anti-inflammatory properties [96]. Various studies surveyed in štefulj et al. [96] implicate LXR in the pathogenesis, modulation, and therapy of AD.

Further potential association between PAPP-A and AD can be seen through study of patients suffering form type-2 diabetes. It has been shown [97] that patients suffering from type-2 diabetes also have significantly increased level of PAPP-A. Akter et al. [98] showed the potentially shared pathology of type-2 diabetes and AD, where some research (e.g. [99]), shows high influence of type-2 diabetes on the potential development of AD. Also, one study performed on mice [100] suggested that changes in the brain during AD can potentially cause diabetes.

## Conclusion

The association of PAPP-A (previously known as pappalysin-1) with cognitive status is probably the most intriguing and novel finding of this study, as it has been scarcely investigated in this context.

PAPP-A was detected as a significant attribute in differentiating between CN, MCI and AD subjects [78] through use of different supervised machine learning algorithms. It has also been shown that it is significant in predicting the progression from MCI to AD, though none of the used subsets of attributes provided adequate predictions of progression between these two classes. High association of PAPP-A with depressive symptoms has already been demonstrated [93] by using the ensemble machine learning algorithm of Random Forests.

In our work, we detected important correlation between the attribute PAPP-A and the cognitive test CDRSB. By applying the newly developed constraint-based extensions of the CLUS-RM algorithm, we detected a larger number of attributes with statistically significant correlation with PAPP-A. In addition to CDRSB, we observed more clinical tests, such as MMSE and ADAS13, with statistically significant correlations with PAPP-A. Interesting and significant correlations were also observed with the biological attributes: volume of the Fusiform gyrus and volume of the Entorhinal cortex both known as being associated with AD [90, 91] with the volume of Entorhinal cortex being significantly reduced even in the mild case of AD [91].

The high importance of our finding lies in the fact that PAPP-A is a metalloproteinase, already known to cleave insulin-like growth factor (IGF) binding proteins (IGFBPs). Perhaps even more importantly, since it also shares similar substrates with the A Disintegrin and Metalloproteinase (ADAM) family of enzymes (the main group of enzymes that act as $\alpha$-secretase to physiologically cleave the amyloid precursor protein (APP) in the so-called non-amyloidogenic pathway [101]), it could be directly involved in the metabolism of the amyloid precursor protein (APP) in the very early stages of AD. Based on the above, the role of PAPP-A in AD should be investigated in greater details.

PLOS ONE | https://doi.org/10.1371/journal.pone.0187364  October 31, 2017                                28 / 35

## Supporting information

**S1 File. Dataset $D_1$ attribute structure.**
(TXT)

**S2 File. Dataset $D_2$ attribute structure.**
(TXT)

**S3 File. Dataset $D_3$ attribute structure.**
(TXT)

**S4 File. Redescriptions obtained on $D_1$ with support in [5, 10] interval.**
(TXT)

**S5 File. Redescriptions obtained on $D_1$ with support in [11, 39] interval.**
(TXT)

**S6 File. Redescriptions obtained on $D_1$ with support in [40, 99] interval.**
(TXT)

**S7 File. Redescriptions obtained on $D_1$ with support in [100, 820] interval.**
(TXT)

**S8 File. Redescriptions obtained on $D_2$ with support in [5, 10] interval.**
(TXT)

**S9 File. Redescriptions obtained on $D_2$ with support in [11, 39] interval.**
(TXT)

**S10 File. Redescriptions obtained on $D_2$ with support in [40, 99] interval.**
(TXT)

**S11 File. Redescriptions obtained on $D_2$ with support in [100, 470] interval.**
(TXT)

**S12 File. Redescriptions obtained on $D_3$ with support in [5, 10] interval.**
(TXT)

**S13 File. Redescriptions obtained on $D_3$ with support in [11, 39] interval.**
(TXT)

**S14 File. Redescriptions obtained on $D_3$ with support in [40, 99] interval.**
(TXT)

**S15 File. Redescriptions obtained on $D_3$ with support in [100, 420] interval.**
(TXT)

**S16 File. Redescriptions obtained on $D_3$ by using constraint-based redescription mining with support larger than 100 subjects.**
(TXT)

**S17 File. Motivation and explanation of statistical tests used in this work.**
(PDF)

**S18 File. Pseudocode of the CLUS-RM algorithm that can use conjunction, negation and disjunction logical operator in redescription query construction and explanation of introduced constraint-based redescription mining extensions.**
(PDF)

## Acknowledgments

Data used in preparation of this article were obtained from the Alzheimer's Disease Neuroimaging Initiative (ADNI) database (http://adni.loni.usc.edu). As such, the investigators within the ADNI contributed to the design and implementation of ADNI and/or provided data, but did not participate in analysis or writing of this report. A complete listing of ADNI investigators can be found at: http://adni.loni.usc.edu/wp-content/uploads/how_to_apply/ADNI_Acknowledgement_List.pdf.

## Author Contributions

**Conceptualization:** Matej Mihelčić, Nada Lavrač, Sašo Džeroski, Tomislav Šmuc.

**Formal analysis:** Matej Mihelčić.

**Funding acquisition:** Goran Šimić, Nada Lavrač, Sašo Džeroski, Tomislav Šmuc.

**Investigation:** Matej Mihelčić, Goran Šimić, Tomislav Šmuc.

**Methodology:** Matej Mihelčić, Tomislav Šmuc.

**Software:** Matej Mihelčić.

**Supervision:** Nada Lavrač, Tomislav Šmuc.

**Validation:** Goran Šimić.

**Visualization:** Matej Mihelčić.

**Writing – original draft:** Matej Mihelčić, Goran Šimić, Mirjana Babić Leko, Nada Lavrač, Sašo Džeroski, Tomislav Šmuc.

**Writing – review & editing:** Matej Mihelčić, Goran Šimić, Mirjana Babić Leko, Nada Lavrač, Sašo Džeroski, Tomislav Šmuc.